\documentclass[review,numbers,sort&compress]{elsarticle}

\usepackage{lineno,hyperref}
\usepackage{hyperref}
\usepackage{amssymb,amsmath}
\usepackage{wrapfig}
\usepackage{caption}
\usepackage[labelformat=simple,textfont=normalfont,singlelinecheck=off,justification=raggedright]{subcaption}

\usepackage{tabularx}
\usepackage{cleveref}

\usepackage[usenames]{color}

\newcommand{\RJHrevise}[1]{\textcolor{black}{#1}}
\usepackage[normalem]{ulem}

\newcommand{\Peclet}[1]{\text{P\'eclet}}
\newcommand{\Pe}{\mathrm{Pe}}

\journal{Journal of Non-Newtonian Fluid Mechanics}

\begin{document}

\begin{frontmatter}

\title{Constitutive modeling of dilute wormlike micelle solutions: shear-induced structure and transient dynamics}

\author{Richard J. Hommel}

\author{Michael D. Graham\corref{mycorrespondingauthor}}
\cortext[mycorrespondingauthor]{Corresponding author}
\ead{mdgraham@wisc.edu}

\address{Department of Chemical and Biological Engineering, University of Wisconsin - Madison, Madison, Wisconsin 53706, USA}

\begin{abstract}
We present a reformulation of the `reactive rod model' (RRM) of Dutta and Graham [Dutta, Sarit and Graham, Michael D., \textit{JNNFM} 251 (2018)], a constitutive model for describing the behavior of dilute wormlike micelle solutions. The RRM treats wormlike micelle solutions as dilute suspensions of rigid Brownian rods undergoing reversible scission and growth in flow. Evolution equations for micelle orientation and stress contribution are coupled to a kinetic reaction equation for a collective micelle length, producing dynamic variations in the length and rotational diffusivity of the rods. This model has previously shown success in capturing many critical steady-state rheological features of dilute wormlike micelle solutions, particularly shear-thickening and -thinning, non-zero normal stress differences, and a reentrant shear stress-shear rate curve, and could fit a variety of steady state experimental data. The present work improves on this framework, which showed difficulty in capturing transient dynamics and high-shear behavior, by reformulating the kinetic equation for micelle growth on a more microstructural (though still highly idealized) basis. In particular, we allow for micelle growth associated with strong alignment of rods and breakage due to tensile stresses along the micelles. This new formulation captures both steady and transient shear rheology in good agreement with experiments. We also find good agreement with available steady state extensional rheology.
\end{abstract}


\end{frontmatter}


\section{Introduction}
\label{sec:Introduction}
Surfactants are amphiphilic molecules consisting of bulky hydrophilic head groups bonded to long-chain hydrophobic tails; beyond some concentration, the critical micelle concentration (CMC), surfactants self-assemble into aggregate structures whose geometry is dictated by the size, shape, and chemistry of the surfactant molecules as well as temperature and the salinity of the solution \cite{Israelachvili2011,Oelschlaeger2010,Lerouge2009,Cates2006}. Among these aggregate structures are spherical and wormlike or rodlike micelles, as well as vesicles and bilayers. At sufficiently high concentrations, wormlike micelles can entangle and form large-scale networks and branched structures, transitioning into a highly viscoelastic gel-like phase. The application of an external field (e.g. flow), or increased temperature, can disrupt these networks and force micelles into distinct structures \cite{Raghavan2001,Keller1998,Wunderlich1987}.
 
Herein we focus on dilute surfactant solutions that form wormlike micelles (WLMs); these solutions have been shown to exhibit remarkable flow dynamics ranging from pronounced shear-thickening and -thinning regimes to shear-induced structure (SIS) formation, as well as numerous instabilities in shear and extensional flows \cite{Perge2014,Wu2018,Mohammadigoushki2017,Fardin2012,Bhardwaj2007}. The addition of small amounts of wormlike micelle-forming surfactants to turbulent flows can produce up to an $80-90\%$ reduction in turbulent drag, which in some cases can exceed the drag-reducing capabilities of the most widely used polymer solutions \cite{Zakin2017,Virk1975,Zakin1996}. Moreover, the self-assembling capabilities of these micellar solutions, through which surfactants are able to reassemble into aggregate structures following mechanical deformations, can also be exploited to overcome the well-known shear-induced breakdown of polymer chains in the pumping of turbulent flows. There has even been recent interest in leveraging the shear-thickening and -thinning properties of dilute surfactant solutions to form less environmentally destructive carrier fluids in oil recovery operations \cite{Yang2002}. The ubiquity of surfactant solutions, including widespread use in household and commercial products, has motivated numerous studies over recent decades into understanding the rheology, dynamics, and flow behavior of these solutions.

Experiments have demonstrated that the behavior of wormlike micelle solutions drastically changes with concentration \cite{Yang2002}. In the upper semi-dilute and concentrated regimes, WLM solutions typically show shear-thinning behavior reminiscent of long-chain polymer solutions. These higher concentration solutions also tend to exhibit, under appropriate conditions, the well-known shear-banding (or gradient banding) instability. This instability is characterized by the development of a macroscopically ``banded'' flow in which fluid separates into two distinct regions of equal shear stress but each supporting a unique shear rate \cite{Dhont2008,Olmsted1999}. The separation of these two regions is often observable through differences in turbidity and/or birefringence. There has been extensive theoretical and experimental treatment of this instability (see \cite{Dhont2008,Olmsted2008} for a comprehensive review).

In the dilute and lower semi-dilute regimes, WLM solutions exhibit pronounced shear-thickening behavior associated with the formation of SISs; at high deformation rates this thickening gives way to stark shear-thinning induced by the breakdown of SISs and strong orientation of the micelles \cite{Rojas2008,Berret1998}. In the dilute regime, SISs take the form of elongated wormlike micelles that can be several times longer than the equilibrium micelle length, in some cases yielding micelles with lengths on the order of microns \cite{Lerouge2009,Keller1998}. In addition to this shear-thickening and -thinning behavior, dilute WLM solutions can display reentrant, or multivalued, flow curves (shear stress vs. shear rate); the existence of a reentrant flow curve is a necessary condition for a relatively unique vorticity banding instability \cite{Herle2007}. In contrast to the shear-banding instability of higher concentrations, vorticity banding requires that a single shear rate be able to support multiple shear stresses.  In circular Couette flow, this instability manifests as stacked ``bands'' along the vorticity axis, where adjacent bands support distinct shear stresses. Again, similar to gradient banding, these bands can often be visualized by differences in turbidity and birefringence \cite{Dhont2008,Herle2007}.


There have been a number of theoretical treatments and models put forth to explain and predict the behavior and dynamics of WLM solutions. The proposed models can be loosely lumped into three categories: population balance models, coupled fluidity models, and microstructural models. One of the earliest treatments of WLM solutions was by Cates and Turner \cite{CatesTurner1990,Turner1992} with the aim of generating a population balance model that accounted for the different stress relaxation mechanisms associated with wormlike micelles, namely micelle scission and rotational diffusion. This work led to the development of an evolution equation for the probability distribution function for a micelle of given length and orientation by considering a reversible kinetic reaction scheme: a given micelle can rupture at any point along its length to form two shorter micelles, while two collinear micelles can also fuse into a single, larger micelle. The model assumes that rods must be strongly aligned for fusion to occur in order to avoid the large energetic penalty associated with bent micelles; this collinearity assumption results in a positive feedback mechanism owing to the decreased angular mobility of longer rods, and this feedback can incur a gelation transition in which micelle length diverges sharply. Further, the competition between relaxation mechanisms yields two distinct limits: a fast-breaking (scission-dominated) limit and a slow-breaking (rotation-dominated) limit.

The Cates and Turner model has shown good agreement with experiments, however, the presence of a continuous spectrum of lengths in this model greatly restricts both its use in studying more complex flows and its incorporation into fluid-dynamical studies. Nevertheless, this model provides a strong mechanistic basis for understanding the dynamics of WLM solutions, and has been used as a foundation for other WLM models both in the dilute and concentrated regimes \cite{Vasquez2007,Dutta2018}. \RJHrevise{We omit a thorough review of the remaining population balance type models as these have been developed primarily for understanding the behavior of concentrated WLM solutions, and thus fall outside the scope of the present work. Readers interested in the modeling and dynamics of concentrated WLM solutions are directed to the Vazquez-Cook-McKinley (VCM) model \cite{Vasquez2007,Zhou2014}, the simplified tube approximation for rapid-breaking micelles (STARM) model by Peterson and Cates \cite{Peterson2020}, and Brownian dynamics simulations of the VCM and reformulated VCM models by Adams and coworkers \cite{Adams2018}.}

We now turn our attention back to models for dilute WLM solutions, and in particular to coupled fluidity models; as the name suggests these models couple well-known and well-studied models for general viscoelastic fluids (e.g. Oldroyd-B, FENE-P, Giesekus) to a fluidity (inverse viscosity) evolution equation originally proposed by Fredrickson for studying thixotropic systems \cite{Fredrickson1970}. The fluidity equation accounts for spontaneous `buildup' and shear-induced `breakdown' of microstructure in the fluid. By coupling the fluidity equation to the Oldroyd-B equation, Bautista et al. \cite{Bautista1999} have formed the BMP model, which is able to predict both shear-thickening and shear-thinning behavior. More recently, Manero et al. \cite{Manero2007} developed the generalized BMP model, where the model's kinetic parameters are taken as functions of the second invariants of both deformation rate and stress tensors, expanding the model to allow for reentrant behavior. This model has also been used successfully by Lanz\`azuri et al. \cite{Landazuri2016} to predict the steady and transient behavior of CTAB and CTAVB solutions and has shown good agreement with experimental results. However, the BMP model lacks a clear connection between model parameters and the underlying microstructural dynamics of WLM solutions and the use of the Oldroyd-B equation presents difficulties in extensional flows due to a divergent extensional viscosity \cite{Boek2005}.  


More recently, Tamano et al. \cite{Tamano2020} have taken inspiration from the BMP model and have coupled a fluidity equation to the Giesekus and FENE-P models to form the f-Giesekus and f-FENE-P models, respectively. This formulation results in four dimensionless parameters accounting for micelle breakdown and build-up (e.g. elongation) timescales, infinite and zero-shear viscosity ratios, and maximum extensibility of micelles. Using this formulation the authors simultaneously capture shear-thickening and -thinning, which is not possible using either of the pure Giesekus and FENE-P models. Further, the transient behavior of this model, in particular in startup of steady shear flow, demonstrates a stress overshoot that is similar to experimental observations of WLM solutions. The f-FENE-P model does suffer from an inability to predict nonzero second normal stress differences, an artifact of the original FENE-P model's inability to do so, and there is some difficulty in predicting the steepness of viscosity vs.~shear rate data as well as the magnitude of shear-thickening seen in experiments. Notably, however, this formulation is particularly well-suited for implementation in direct numerical simulations (DNS) given that the majority of simulations for viscoelastic polymer solutions are already formulated on the FENE-P and Giesekus constitutive equations.

The present work concerns the last class of wormlike micelle models -- microstructural models 
 -- and in particular focuses on a reformulation of the reactive rod model (RRM), a tensor constitutive model proposed by Dutta and Graham \cite{Dutta2018}. The RRM takes a phenomenological, but highly microstructurally-motivated, approach  to modeling WLMs and treats dilute wormlike micelle solutions as suspensions of rigid Brownian rods able to undergo reversible growth and scission in flow. The model couples evolution equations governing the ensemble average orientation of rods and micelle stress contribution to an evolution equation for the collective length of micelles, where micelle number density and length are constrained by conservation of surfactant molecules. We provide more details on the modeling framework of the RRM in \cref{sec:ModelDescription}, but note that it follows mechanisms proposed by Turner and Cates \cite{Turner1992}, namely growth due to flow-induced alignment of micelles and spontaneous scission along the lengths of micelles. The RRM has shown success in capturing experimental observations of dilute WLM solutions, and is able to capture flow curve multiplicity, a necessary condition for vorticity banding, as well as nonzero normal stress differences. While this model provides a proof of principle that the modeling structure of rigid Brownian rods is rich enough to capture many key features associated with dilute WLM solutions, it does have difficulty in predicting transient flow and high-shear results. Further, the nature of the proposed length equation lacks clear physical insights that, if present, would allow for more precise understanding of rheological behavior. The motivation of the present work is to further develop the RRM by developing a length evolution equation based more closely on WLM dynamics in flow, with the aim of forming a model that is both accurate in predicting WLM rheology and tractable enough for implementation in DNS and in the study of more complex flows.


\section{Model description}
\label{sec:ModelDescription}

The complete derivation of the \textit{reactive rod model} (RRM) is described in \cite{Dutta2018}. As discussed above, this modeling framework takes inspiration from theoretical treatments by Cates and Turner \cite{CatesTurner1990}.
In summary, dilute wormlike micelle solutions are treated as suspensions of rigid Brownian rods undergoing reversible scission and growth. Rods fuse end-to-end (reducing the energetic penalty associated with the micellar end caps), but only when they are highly aligned -- otherwise the energy penalty arising from forming a long but bent micelle is too large for fusion to take place \cite{Larson1999,CatesTurner1990}. 
 The application of flow tends to align the rods. This alignment is balanced by rotational diffusivity of the rods acting to return the suspension to isotropy. The fundamental assertion of the RRM is that rods are able to react (fuse) in flow to form longer rods, and that the reaction rate increases with increasing rod alignment. Consequently, a positive feedback mechanism exists between rod growth and alignment owing to the smaller rotational diffusivity of longer rods. It is assumed that rod growth is countered by hydrodynamic stresses acting along the lengths of the rods and these stresses, which increase with increasing length, can induce breakage events into shorter rods. 
 
\subsection{Rigid Brownian rods}
The underlying theory of a non-reactive suspension of rigid Brownian rods is given in both \cite{Dutta2018,Doi1986}, which we briefly review before delving into the complete RRM. We begin with a uniform collection of rods with length $L_0$, radius $b$, and number density $n_0$ suspended in a Newtonian solvent with viscosity $\eta_s$. The orientation of a single rod is described by the unit director vector $\boldsymbol{u}$. The suspension is subjected to an arbitrary, homogeneous flow with velocity $\boldsymbol{v}$ and transpose velocity gradient $\boldsymbol{K} = \boldsymbol{\nabla}\boldsymbol{v}^\top$. The orientation tensor $\boldsymbol{S}$ describes the average collective orientation of the suspension and is given by the second moment of $\boldsymbol{u}$
\begin{equation}
    \boldsymbol{S} = \left\langle\boldsymbol{uu}\right\rangle = \int\boldsymbol{uu}\psi\mathrm{d}\boldsymbol{u},
\end{equation}
where $\psi$ is the probability distribution function of $\boldsymbol{u}$. The time evolution of $\boldsymbol{S}$ in a homogeneous flow can be found by multiplying the rotational Smoluchowski equation
\begin{equation}
	\frac{\partial \psi}{\partial t} = D_{r,0} \mathcal{R}^2\psi - \mathcal{R} \cdot \left(\boldsymbol{u} \times \boldsymbol{K}\cdot \boldsymbol{u}\psi\right),
	\end{equation}
where $\mathcal{R} \equiv \boldsymbol{u} \times \frac{\partial}{\partial \boldsymbol{u}}$ and $D_{r,0}$ is the rotational diffusion coefficient of a rod, by $\boldsymbol{S}$ and integrating over $\boldsymbol{u}$ \cite{Doi1986}. We then have for the time evolution of $\boldsymbol{S}$
\begin{equation}
    \frac{\mathrm{d}\boldsymbol{S}}{\mathrm{d}t} = -6D_{r,0}\left(\boldsymbol{S} - \frac{1}{3}\boldsymbol{I}\right) + \boldsymbol{K}\cdot\boldsymbol{S}^\top + \boldsymbol{S}\cdot\boldsymbol{K}^\top - 2\boldsymbol{K}:\left\langle\boldsymbol{uuuu}\right\rangle,
    \label{eqn:Orientation1}
\end{equation}
where $\boldsymbol{I}$ is the unit tensor and the double dot product is defined as $\boldsymbol{A}:\boldsymbol{B} = \text{Tr}(\boldsymbol{A}\cdot\boldsymbol{B}^\top)$.

The total stress of the suspension is given by the sum of the solvent $\boldsymbol{\tau}_s$ and micellar $\boldsymbol{\tau}_m$ contributions
\begin{equation}
    \boldsymbol{\tau} = \boldsymbol{\tau}_s + \boldsymbol{\tau}_m,
\end{equation}
where
\begin{equation}
    \boldsymbol{\tau}_s = 2\eta_s\boldsymbol{D}
\end{equation}
is the Newtonian solvent contribution with rate of deformation tensor $\boldsymbol{D} = \frac{1}{2}(\boldsymbol{K} + \boldsymbol{K}^\top)$ and
\begin{equation}
    \boldsymbol{\tau}_m = 3n_0k_BT\left(\boldsymbol{S}-\frac{1}{3}\boldsymbol{I}\right) + \frac{n_0k_BT}{2D_{r,0}}\boldsymbol{K}:\left\langle\boldsymbol{uuuu}\right\rangle
    \label{eqn:StressP}
\end{equation}
is the additional stress due to the presence of rods. Here, $k_B$ is the Boltzmann constant and $T$ is the temperature. \Cref{eqn:Orientation1,eqn:StressP} notably contain the fourth moment $\langle \boldsymbol{uuuu} \rangle$, an evolution equation for which depends on the sixth moment of $\boldsymbol{u}$, which in turn depends on higher moments. To proceed analytically, it is then necessary to supply a closure approximation for the product $\boldsymbol{K}:\left\langle\boldsymbol{uuuu}\right\rangle$. While numerous approximations are possible (see, for example: \cite{Doi1986,Dhont2006,Forest2003}), we use an approximation from Dhont and Briels \cite{Dhont2003} that interpolates between exact expressions in the limits of isotropy (equilibrium) and complete alignment:
\begin{multline}
    \boldsymbol{K}:\left\langle\boldsymbol{uuuu}\right\rangle \approx \frac{1}{5}[\boldsymbol{S}\cdot\boldsymbol{D}+\boldsymbol{D}\cdot\boldsymbol{S}-\boldsymbol{S}\cdot\boldsymbol{S}\cdot\boldsymbol{D}-\boldsymbol{D}\cdot\boldsymbol{S}\cdot\boldsymbol{S}+ ... \\ 2\boldsymbol{S}\cdot\boldsymbol{D}\cdot\boldsymbol{S}+3(\boldsymbol{S}:\boldsymbol{D})\boldsymbol{S}].
    \label{eqn:Closure}
\end{multline}
\subsection{Reactive Brownian rods}
As discussed above, a key feature of the RRM is that it allows micelles, modeled as rigid rods, to undergo reversible scission and growth by allowing the collective length and number density of the suspension to be dynamic properties that evolve with time and flow. Consider a suspension of rods at equilibrium initially with number density $n_0$ and representative length $L_0$; in practice the length of a wormlike micelle suspension follows an exponential distribution \cite{Larson1999}, but to make analytical progress it is assumed that a single, representative length is able to suitably characterize the system. The radius $b$ of the rods is taken to be constant. The evolution of  length $L$ and number density $n$ are constrained at all times by the surfactant mass balance
 \begin{equation}
nL = n_0L_0. 	\label{eq:surfactantbalance}
 \end{equation}
Everywhere below, $n$ will be determined by this equation. 

The rotational diffusion constant for a rod of length $L_0$ and radius $b$ is given by \cite{Doi1986}
\begin{equation}
    D_{r,0} = \frac{3k_BT}{\pi\eta_sL_0^3}\ln\left(\frac{L_0}{2b}\right).
\end{equation}
In the RRM, the constant rotational diffusion coefficient of the simple rigid rod model is replaced by the length-dependent coefficient
\begin{equation}
    D_r = \frac{D_{r,0}}{L^{*3}}\left(\frac{\ln L^* + m}{m}\right),
    \label{eqn:DiffusionCoeff}
\end{equation}
where $L^* = L/L_0$ is the dimensionless micelle length and $m = \ln[L_0/(2b)]$ is a constant related to the initial aspect ratio of the rods. Substituting \cref{eqn:DiffusionCoeff} into \cref{eqn:Orientation1,eqn:StressP}, we find
\begin{equation}
    \frac{\mathrm{d}\boldsymbol{S}}{\mathrm{d}t} = -6D_{r}\left(\boldsymbol{S} - \frac{1}{3}\boldsymbol{I}\right) + \boldsymbol{K}\cdot\boldsymbol{S}^\top + \boldsymbol{S}\cdot\boldsymbol{K}^\top - 2\boldsymbol{K}:\left\langle\boldsymbol{uuuu}\right\rangle
    \label{eqn:RRM_Orientation_OG}
\end{equation}
and
\begin{equation}
    \boldsymbol{\tau}_m = 3nk_BT\left(\boldsymbol{S}-\frac{1}{3}\boldsymbol{I}\right) + \frac{nk_BT}{2D_{r}}\boldsymbol{K}:\left\langle\boldsymbol{uuuu}\right\rangle.
    \label{eqn:RRM_Stress_OG}
\end{equation}

The rod orientation of the suspension is tracked by introducing a  scalar orientational order parameter
\begin{equation}
    \widehat{S} = \sqrt{\frac{3}{2}\widehat{\boldsymbol{S}}:\widehat{\boldsymbol{S}}},
\end{equation}
where $\widehat{\boldsymbol{S}} = \boldsymbol{S} - \frac{1}{3}\boldsymbol{I}$ is the traceless part of $\boldsymbol{S}$. This order parameter varies between $\widehat{S} = 0$ for isotropic rods and $\widehat{S} = 1$ for perfectly aligned rods. The orientation and stress equations are rendered dimensionless by introducing a nondimensional time $t^* = D_{r,0}t$ and P\'eclet  number $\mathrm{Pe} = \dot{\gamma}/D_{r,0}$ in shear flow and $\mathrm{Pe} = \dot{\epsilon}/D_{r,0}$ in extensional flow. Note that the description and equations above are valid for both the original RRM and the reformulation (RRM-R), the only variation between the two models is in the length evolution equation, discussed below. 

\subsection{Reactive rod model (RRM)}
To allow for variability of rod length the original RRM introduces a length evolution equation of the form
\begin{equation}
    \frac{\mathrm{d}L^*}{\mathrm{d}t^*} = R_a + R_s,
    \label{eqn:Length1}
\end{equation}
where $R_a$ represents alignment-induced growth of micelles and $R_s$ collectively represents spontaneous fusion and breakdown of micelles. The model assumes that alignment-induced growth increases linearly with rod alignment, while spontaneous growth and breakage are assumed to be proportional to the instantaneous deviation from equilibrium micelle length, $L_0$. The RRM introduces the idea of a maximum length $L^*_{\text{max}}$ beyond which micelles are broken down by hydrodynamic stresses, and since these stresses increase with increasing deformation rates, $L^*_{\text{max}}$ must decrease with increasing $\mathrm{Pe}$. Moreover, the breakdown rate is assumed to increase without bound as $L$ approaches $L^*_{\text{max}}$, suggesting a FENE-like form for $R_s$. Using these assumptions, the RRM proposes a complete length evolution equation of the form
\begin{equation}
    \frac{\mathrm{d}L^*}{\mathrm{d}t^*} = \frac{\lambda}{1-\left(\frac{L^*}{\alpha + \frac{\beta}{\mathrm{Pe}}}\right)^2}(1-L^*) + k\widehat{S},
    \label{eqn:RRM_Length_OG}
\end{equation}
where $\lambda$, $k$, $\alpha$, and $\beta$ are model parameters. \Cref{eqn:Closure,eqn:RRM_Orientation_OG,eqn:RRM_Stress_OG,eqn:RRM_Length_OG} now form a closed set of equations describing the orientation, stress, and length of a suspension of reactive Brownian rods. 

\subsection{Reformulated reactive rod model (RRM-R)}
The original formulation of the RRM does an excellent job in capturing some of the most commonly seen phenomena in dilute wormlike micelle rheology, namely shear- and extensional-thickening and -thinning, reentrant (i.e. multivalued) shear stress, and non-zero normal stress differences. It also makes progress in putting forth a tractable constitutive model that has potential for implementation in CFD simulations. In this way, the original RRM acts as a proof-of-principle, indicating that the coupling of orientation, stress, and length evolution equations has great potential in the modeling of dilute wormlike micelle rheology. 

However, the original formulation of the RRM has difficulty in accurately capturing phenomena associated with transient flows (e.g. inception of steady shear), which is a significant obstacle for future implementation in CFD studies. \RJHrevise{This difficulty  arises from the nature of length evolution equation (\cref{eqn:RRM_Length_OG}), which notably depends only on \Peclet{} number (which is fixed for a given flow) and not on stress (which depends on flow type and time). Importantly, this lack of stress dependence prevents the original RRM from fully distinguishing between shear and extensional flows, or between steady and transient flows.} Further, while the length equation in the original RRM does incorporate experimental insights about micellar rheology, such as alignment-induced growth and spontaneous growth and breakdown, there is not a clear physical grounding underlying the structure of the proposed equation. To address this, we present a new length evolution equation that considers in greater depth the microscopic mechanisms underlying micellar reversible scission dynamics. This new formulation incorporates mechanistic parameters that represent distinct microstructural phenomena, thus facilitating more direct comparisons with surfactant rheology and chemistry. It must be acknowledged that this new model still does not attempt to capture the distribution of micelle lengths, nor does it encompass other effect such as formation of branched or laterally-associating micelles.

We begin our reformulation with a similar approach to the original RRM, assuming a length evolution equation that balances growth and breakdown of micelles
\begin{equation}
    \frac{\mathrm{d}L}{\mathrm{d}t} = R_g + R_b,
    \label{eqn:LengthNew}
\end{equation}
where $R_g \geq 0$ is the rate of micelle growth and $R_b \leq 0$ is the rate of micelle breakdown. 
\subsubsection{Micelle growth rate}
Beginning with $R_g$, we assume that the rate of growth consists of two distinct types: spontaneous growth, $R_{g,s}$, and alignment-induced growth, $R_{g,a}$
\begin{equation}
    R_g = R_{g,s} + R_{g,a}.
\end{equation}
Spontaneous growth must occur both in the presence and absence of flow and, as we will address later, must balance with spontaneous breakage in a quiescent suspension to maintain equilibrium. Spontaneous growth, $R_{g,s}$, which occurs when the ends of two micelles randomly collide, must increase with number density and thus we propose a form 
\begin{equation}
    R_{g,s} = n^2k_{g0},
\end{equation}
where $n$ is number density and $k_{g0} \; [\mathrm{m}^7 \mathrm{s}^{-1}]$ is a spontaneous growth rate coefficient. The choice of a quadratic dependence on number density is due to the bimolecular nature of a collision event. We note that because we limit our framework to dilute solutions, interactions between micelles, and therefore spontaneous growth events, should occur infrequently.

Now turning our attention to alignment-induced growth, a feature of both Cates and Turner's work as well as the original RRM, we assume that this type of growth must depend on both the alignment and collision frequency of rods. We capture this idea with a functional form
\begin{equation}
    R_{g,a} = f(\widehat{S})g(\nu),
\end{equation}
where $\nu$ is collision frequency. We assume a separable form here to make analytical progress, though this may not be the case. Rod alignment is captured by the orientational order parameter $\widehat{S}$, and the bimolecular nature of a combination event suggests a quadratic dependence on order
\begin{equation}
	f(\widehat{S}) = k_{ga,1}{\widehat{S}}^2,
\end{equation}
where $k_{ga,1}$ is a kinetic growth parameter. 


There are two possible mechanisms for collisions -- diffusion and flow. A simple approximation for collision frequency due to flow is $\nu = n^2||\boldsymbol{D}||$. Here, $\nu$ depends on $\boldsymbol{D}$ rather than $\boldsymbol{K}$ because strain and the relative motion of micelles, as opposed to rigid rotation, produces collisions. Again, the bimolecular nature of growth suggests a quadratic dependence on number density. Ignoring variation in number density with flow rate we thus approximate that the collision frequency due to flow scales linearly with \Peclet{} number.  

Now turning our attention towards collisions due to diffusion, we can show that translational diffusivity alone, and not strain induced by the flow, is sufficient to induce end-to-end collisions. The mean squared displacement $\Delta r^2$ behaves as 
\begin{equation}
    \left\langle \Delta r^2 \right\rangle = 6 D_\mathrm{eff} t,
    \label{eqn:meanSqr}
\end{equation}
where $D_\mathrm{eff}=(D_\parallel+2D_\perp)/3$ is the effective translational diffusion coefficient for the rod, and $D_\parallel$ and $D_\perp\approx D_\parallel/2$ are the diffusivities parallel and perpendicular to the rod axis, respectively  \cite{Graham:2018ty}. The parallel diffusivity is given by 
\begin{equation}
    D_\parallel = \frac{k_BT\ln(L/b)}{2\pi \eta_sL}
    \label{eqn:parallelDiffusion}
\end{equation}
\cite{Doi1986}.
For a typical wormlike micelle of length $L \sim\mathcal{O}(10 \;\mathrm{nm})$ and radius $b \sim\mathcal{O}(1\; \mathrm{nm})$ in an aqueous solution at room temperature, we find from \cref{eqn:parallelDiffusion} that $D_\parallel \sim \mathcal{O}(10^{-11}\;\mathrm{m}^2 \mathrm{s}^{-1}) $. 


\RJHrevise{We estimate the average distance between between rods as $x_\text{avg} \sim n^{-1/3}$. We consider a dilute surfactant solution with concentration $\sim\mathcal{O}(1\;\mathrm{mM})$ and an estimated $\sim\mathcal{O}(10^2)$ surfactant molecules per rod, giving a rod number density of $\sim\mathcal{O}(10^{23}\; \mathrm{rods}/\mathrm{m}^3)$, and thus an average distance between rods of $x_\text{avg} \sim 10^{-8}\; \mathrm{m}$. Using \cref{eqn:meanSqr}, we calculate that end-to-end collision events occur on the order of $\sim\mathcal{O}(10^{-5}\;\mathrm{s})$. This fact that diffusion-driven collision events occur on such a short timescale indicates that diffusion alone is enough to account for collision-induced growth. Of course, at sufficiently high shear rates (e.g. $\gtrsim 10^4$ s$^{-1}$), such as in turbulent flows, convective and diffusive timescales will become comparable and this approximation will no longer hold; however, the majority of rheological experiments occur below this high shear regime. We therefore can discard the convective shear dependence in the alignment-induce growth term, leaving us with $g(\nu) = k_{ga,2}n^2$, and our overall growth rate is}

\begin{equation}
    R_g = n^2k_{g0} + k_{ga}n^2{\widehat{S}}^2,
    \label{eqn:growthFinal}
\end{equation}
where $k_{ga} \; [m^7 s^{-1}]$ is an overall alignment-induced growth coefficient.

\subsubsection{Micelle breakage rate}
We now turn our attention to the micellar breakage rate $R_b$. Similar to the growth rate, we consider two distinct types of breakage: spontaneous breakage, $R_{b,s}$, and tension-induced breakage, $R_{b,t}$
\begin{equation}
    R_b = R_{b,s} + R_{b,t}.
    \label{eqn:breakageRate}
\end{equation}
Spontaneous breakage, like growth, occurs in both the presence and absence of flow and must increase with rod length. These considerations suggest a spontaneous breakage rate of the form
\begin{equation}
    R_{b,s} = -k_{b0}L,
    \label{eqn:Rbeq}
\end{equation}
where $k_{b0} \; [s^{-1}]$ is a rate constant. At equilibrium spontaneous growth and breakage must balance such that
\begin{equation}
    0 = n_0^2k_{g0} - k_{b0}L_0.
    \label{eqn:eqLength}
\end{equation}
This relation, along with the surfactant mass balance (\cref{eq:surfactantbalance}), allows us to rewrite $k_{g0}$ in terms of $k_{b0}$.

Now considering breakage under flow, we assume that as rods become sufficiently long they will be broken down by hydrodynamic stresses. Although this mechanism was incorporated in the original RRM, we propose here a more physical grounding for the functional form of the breakage term, by specifically taking this rate to be affected by the tensile force $\boldsymbol{T}$ at the midpoint of a rod. This force acts along the direction of the rod, so can be written $\boldsymbol{T}=T\boldsymbol{u}$. Now we note that the stress $\boldsymbol{\tau}_m$ exerted on the fluid by the rods is approximated by 
\begin{equation}
    \boldsymbol{\tau}_m \sim n(\boldsymbol{T}L\boldsymbol{u})=n(TL\boldsymbol{u}\boldsymbol{u}).
\end{equation}
Here the term in parentheses estimates the force dipole exerted by a micelle on the fluid  \cite{Graham:2018ty}.
Taking the dot product with $\boldsymbol{u}$ on both the left and right, noting that $\boldsymbol{u} \cdot \boldsymbol{\tau}_m \cdot \boldsymbol{u} = \boldsymbol{uu} : \boldsymbol{\tau}_m$, and solving for $T$ yields
\begin{equation}
    T \sim \frac{\boldsymbol{uu}:\boldsymbol{\tau}_m}{nL}.
    \label{eqn:forceApprox2}
\end{equation}
We can further simplify this expression by estimating $\boldsymbol{u}\boldsymbol{u}$ as its ensemble average $\boldsymbol{S}$ and applying the surfactant balance $nL = n_0L_0$ to find that
\begin{equation}
    T \sim \frac{\boldsymbol{S}:\boldsymbol{\tau}_m}{n_0L_0}.
    \label{eqn:forceApprox3}
\end{equation}
Viewing the micelle tension as increasing the likelihood that a micelle will overcome the free energy barrier to scission  \cite{Larson1999} motivates the use of an Arrhenius-type breakage rate expression:
\begin{equation}
    R_{b,t} = -k_{bt}\left[\exp\left(\frac{a}{k_BT}\frac{\boldsymbol{S}:\boldsymbol{\tau}_m}{n_0L_0}\right)-1\right],
    \label{eqn:Rbt}
\end{equation}
where $k_{bt} \; [m s^{-1}]$ is a tension-induced breakage coefficient and $a \; [m]$ acts as constant for micelle scission. Note that $Ta$ has units of work, and thus one could view $a$ as the distance the two halves of the micelle need to be pulled apart to break it in half. In our fitting of experimental data, we find $a$ to be on the order of nanometers, which is physically reasonable. \RJHrevise{This term has been structured so that at rest, (i.e.~when $\boldsymbol{S}:\boldsymbol{\tau}_m = 0$), the tension-induced breakage rate vanishes entirely.} Substituting \cref{eqn:Rbeq,eqn:Rbt} into \cref{eqn:breakageRate} gives our overall breakage rate
\begin{equation}
        R_b = -k_{b0}L - k_{bt}\left[\exp\left(\frac{a}{k_BT}\frac{\boldsymbol{S}:\boldsymbol{\tau}_m}{n_0L_0}\right)-1\right]
        \label{eqn:breakageFinal}
\end{equation}

\subsubsection{Overall length evolution}
Substituting \cref{eqn:growthFinal,eqn:breakageFinal} into \cref{eqn:LengthNew} gives our dimensional length evolution equation
\begin{equation}
    \frac{\mathrm{d}L}{\mathrm{d}t} = k_{g0}n^2 + k_{ga}n^2\widehat{S}^2 - k_{b0}L - k_{bt}\left[\exp\left(\frac{a}{k_BT}\frac{\boldsymbol{S}:\boldsymbol{\tau}_m}{n_0L_0}\right)-1\right].
\end{equation}
We can further simplify this expression using \cref{eqn:eqLength} and the surfactant mass balance to obtain
\begin{equation}
    \frac{\mathrm{d}L}{\mathrm{d}t} = k_{b0}\left(\frac{L_0^3}{L^2} - L\right) + k_{ga}(n_0L_0)^2\frac{\widehat{S}^2}{L^2} - k_{bt}\left[\exp\left(\frac{a}{k_BT}\frac{\boldsymbol{S}:\boldsymbol{\tau}_m}{n_0L_0}\right)-1\right].
    \label{eqn:LengthDimensional}
\end{equation}
We render \cref{eqn:LengthDimensional} dimensionless by introducing a nondimensional time $t^* = tD_{r,0}$ and length $L^* = L/L_0$
\begin{equation}
    \frac{\mathrm{d}L^*}{\mathrm{d}t^*} = k_{b0}^*\left(\frac{1}{L^{*2}} - L^*\right) + k_{ga}^*\frac{\widehat{S}^2}{L^{*2}} - k_{bt}^*\left[\exp\left(a^*\boldsymbol{S}:\boldsymbol{\tau}^*_m\right)-1\right],
    \label{eqn:LengthDimensionless}
\end{equation}
where
\begin{equation}
	\boldsymbol{\tau}_m^* = \frac{\boldsymbol{\tau}_m}{n_0k_BT},
\end{equation}
and
\begin{equation}
    k_{b0}^* = \frac{k_{b0}}{D_{r,0}}, \quad k_{ga}^* = \frac{k_{ga}n_0^2}{D_{r,0}L_0}, \quad k_{bt}^* = \frac{k_{bt}}{D_{r,0}L_0}, \quad a^* = \frac{a}{L_0}.
\end{equation}

\vspace{2mm}
We have now introduced four dimensionless groups: $k_{b0}^*$, $k_{ga}^*$, $k_{bt}^*$, and $a^*$. In order, $k_{b0}^*$ represents the ratio of relaxation due to spontaneous breakage and relaxation due to diffusion (i.e. realignment), $k_{ga}^*$ acts as a measure of the ratio of growth due to alignment to diffusion, $k_{bt}^*$ represents the ratio of relaxation due to tension-induced breakage and relaxation due to diffusion, and finally, $a^*$, as noted above, functions as a dimensionless length that must be overcome for tension-induced scission to occur. \Cref{eqn:DiffusionCoeff,eqn:Closure,eqn:RRM_Orientation_OG,eqn:RRM_Stress_OG,eqn:LengthDimensionless} form a closed set of ODEs governing the time-evolution of a dilute wormlike micelle solution modeled as reactive Brownian rods.

\subsection{Shear flow}
For a simple shear velocity profile $\boldsymbol{v} = [\dot{\gamma}y,0,0]^\top$ with \Peclet{} number $\mathrm{Pe} = \dot{\gamma}/D_{r,0}$, substituting our closure relation \cref{eqn:Closure} into the orientation equation \cref{eqn:RRM_Orientation_OG} and supplying \cref{eqn:DiffusionCoeff} to account for variation in the rotational diffusivity gives
\begin{subequations}
    \begin{equation}
        \frac{\partial S_{x x}}{\partial t^{*}}=-\frac{6}{L^{* 3}}\left(\frac{\ln L^{*}+m}{m}\right)\left(S_{x x}-\frac{1}{3}\right)-\frac{2}{5} \operatorname{Pe} S_{x y}\left(-4+4 S_{x x}-S_{y y}\right),
    \end{equation}
    \begin{equation}
        \frac{\partial S_{y y}}{\partial t^{*}} =-\frac{6}{L^{* 3}}\left(\frac{\ln L^{*}+m}{m}\right)\left(S_{y y}-\frac{1}{3}\right)-\frac{2}{5} \mathrm{Pe} S_{x y}\left(1+4 S_{y y}-S_{x x}\right),
    \end{equation}
    \begin{equation}
        \frac{\partial S_{z z}}{\partial t^{*}} =-\frac{6}{L^{* 3}}\left(\frac{\ln L^{*}+m}{m}\right)\left(S_{z z}-\frac{1}{3}\right)-\frac{6}{5} \mathrm{Pe} S_{x y} S_{z z},
    \end{equation}
    \begin{equation}
        \frac{\partial S_{x y}}{\partial t^{*}}=-\frac{6}{L^{* 3}}\left(\frac{\ln L^{*}+m}{m}\right) S_{x y}-\frac{\mathrm{Pe}}{5}\left[6 S_{x y}^{2}+S_{x x}-4S_{y y}-\left(S_{x x}-S_{y y}\right)^{2}\right].
    \end{equation}
    \label[equation]{eqn:ShearOrientation}
\end{subequations}
Likewise we have the components of the stress tensor
\begin{subequations}
    \begin{equation}
       \tau_{m,x x}^*=\frac{3}{L^{*}}\left(S_{x x}-\frac{1}{3}\right)+\frac{m \mathrm{Pe} L^{* 2}}{10\left(\ln L^{*}+m\right)} S_{x y}\left(1+4 S_{x x}-S_{y y}\right),
    \end{equation}
    \begin{equation}
        \tau_{m,y y}^* =\frac{3}{L^{*}}\left(S_{y y}-\frac{1}{3}\right)+\frac{m \mathrm{Pe} L^{* 2}}{10\left(\ln L^{*}+m\right)} S_{x y}\left(1+4 S_{y y}-S_{x x}\right),
    \end{equation}
    \begin{equation}
        \tau_{m,z z}^*=\frac{3}{L^{*}}\left(S_{z z}-\frac{1}{3}\right)+\frac{3m \mathrm{Pe} L^{* 2}}{10\left(\ln L^{*}+m\right)} S_{x y} S_{z z},
    \end{equation}
    \begin{equation}
        \tau_{m,x y}^*=3\frac{S_{x y}}{L^{*}}+\frac{m \mathrm{Pe} L^{* 2}}{20\left(\ln L^{*}+m\right)}\left[6 S_{x y}^{2}+S_{x x}+S_{y y}-\left(S_{x x}-S_{y y}\right)^{2}\right].
    \end{equation}
    \label[equation]{eqn:ShearStress}
\end{subequations}
In simple shear flow our length evolution \cref{eqn:LengthDimensionless} becomes
\begin{gather}
    \frac{\mathrm{d}L^*}{dt^*} = k_{b0}^*\left(\frac{1}{L^{*2}}-L^*\right) + k_{ga}^*\frac{\widehat{S}^2}{L^{*2}}\mathrm{Pe} - k_{bt}^*\left[\exp\left(a^*\boldsymbol{S}:\boldsymbol{\tau}^*_m\right)-1\right],
    \label{eqn:ShearLengthDimless}
\end{gather}
with
\begin{equation}
    \boldsymbol{S}:\boldsymbol{\tau}_m^* = S_{xx}\tau_{m,xx}^* + S_{yy}\tau_{m,yy}^* + S_{zz}\tau_{m,zz}^* + 2S_{xy}\tau_{m,xy}^*,
\end{equation}
and with scalar orientation parameter
\begin{equation}
    \widehat{S}=\left[\frac{3}{2}\left\{\left(S_{x x}-\frac{1}{3}\right)^{2}+\left(S_{y y}-\frac{1}{3}\right)^{2}+\left(S_{z z}-\frac{1}{3}\right)^{2}+2 S_{x y}^{2}\right\}\right]^{\frac{1}{2}},
    \label{eqn:ShearShat}
\end{equation}
where $S_{xx} + S_{yy} + S_{zz} = 1$.

\subsection{Uniaxial extensional flow}
Following similar steps for uniaxial extensional flow with velocity profile $\boldsymbol{v} = [-\dot{\epsilon}x/2,-\dot{\epsilon}y/2,\dot{\epsilon}z]^\top$ and \Peclet{} number $\mathrm{Pe} = \dot{\epsilon}/D_{r,0}$, our orientation equations are
\begin{subequations}
    \begin{equation}
        \frac{\partial S_{x x}}{\partial t^{*}}=-\frac{6}{L^{* 3}}\left(\frac{\ln L^{*}+m}{m}\right)\left(S_{x x}-\frac{1}{3}\right)-\frac{9}{5}\mathrm{Pe} S_{x x} S_{z z},
    \end{equation}
    \begin{equation}
        \frac{\partial S_{z z}}{\partial t^{*}}=-\frac{6}{L^{* 3}}\left(\frac{\ln L^{*}+m}{m}\right)\left(S_{z z}-\frac{1}{3}\right)-\frac{9}{5}\mathrm{Pe}\left(S_{z z}^{2}-S_{z z}\right).
    \end{equation}
\end{subequations}
Likewise we have the components of the stress tensor
\begin{subequations}
    \begin{equation}
        \tau_{m,x x}^*=\frac{3}{L^{*}}\left(S_{x x}-\frac{1}{3}\right)+\frac{m \mathrm{Pe} L^{* 2}}{20\left(\ln L^{*}+m\right)} S_{x x}\left(9 S_{z z}-5\right),
    \end{equation}
    \begin{equation}
        \tau_{m,zz}^*=\frac{3}{L^{*}}\left(S_{z z}-\frac{1}{3}\right)+\frac{m \mathrm{Pe}{L}^{* 2}}{20\left(\ln L^{*}+m\right)}\left(9 S_{z z}^{2}+S_{z z}\right).
    \end{equation}
\end{subequations}
Our length evolution \cref{eqn:LengthDimensionless} in uniaxial extension becomes
\begin{equation}
    \frac{\mathrm{d}L^*}{dt^*} = k_{b0}^*\left(\frac{1}{L^{*2}}-L^*\right) + k_{ga}^*\frac{\sqrt{6}}{2}\frac{\widehat{S}^2}{L^{*2}}\mathrm{Pe} - k_{bt}^*\left[\exp\left(a^*\boldsymbol{S}:\boldsymbol{\tau}_m^*\right)-1\right],
\end{equation}
with
\begin{equation}
    \boldsymbol{S}:\boldsymbol{\tau}_m^* = S_{xx}\tau_{m,xx}^* + S_{yy}\tau_{m,yy}^* +S_{zz}\tau_{m,zz}^*,
\end{equation}
and with scalar orientation parameter
\begin{equation}
    \hat{S}=\left[\frac{3}{2}\left\{\left(S_{x x}-\frac{1}{3}\right)^{2}+\left(S_{y y}-\frac{1}{3}\right)^{2}+\left(S_{z z}-\frac{1}{3}\right)^{2}\right\}\right]^{\frac{1}{2}},
\end{equation}
where symmetry requires $S_{xx} = S_{yy} = (1-S_{zz})/2$.



\section{Results and Discussion}
The organization of this section is as follows: in \cref{sec:SteadyShearParameters} we characterize the effects of the four dimensionless parameters in the RRM-R on steady shear flow and show that by varying the parameters we can capture many key features associated with WLM solution rheology. In \cref{sec:SteadyShearExpr} we fit our model to experimental data reported in literature for steady shear flows and demonstrate that the model is able to predict the behavior of both dilute and semi-dilute WLM solutions, as well as the behavior of both cationic and non-ionic surfactant solutions. We then turn our attention to transient flows in \cref{sec:TransientShear} by characterizing the predictions of the RRM-R in startup of steady shear flow; we further show excellent agreement in simultaneously fitting our model to both steady and transient shear experimental data. Finally, in \cref{sec:SteadyExtensional}, we briefly show that the RRM-R is able to suitably predict the behavior of steady extensional flows of wormlike micelle solutions.  

\subsection{Steady shear: parameter dependence}
\label{sec:SteadyShearParameters}
\begin{figure*}[t]
    \begin{subfigure}{.5\textwidth}
        \centering
        \subcaption{}
        \vspace{-3mm}
        \includegraphics[width=.95\linewidth]{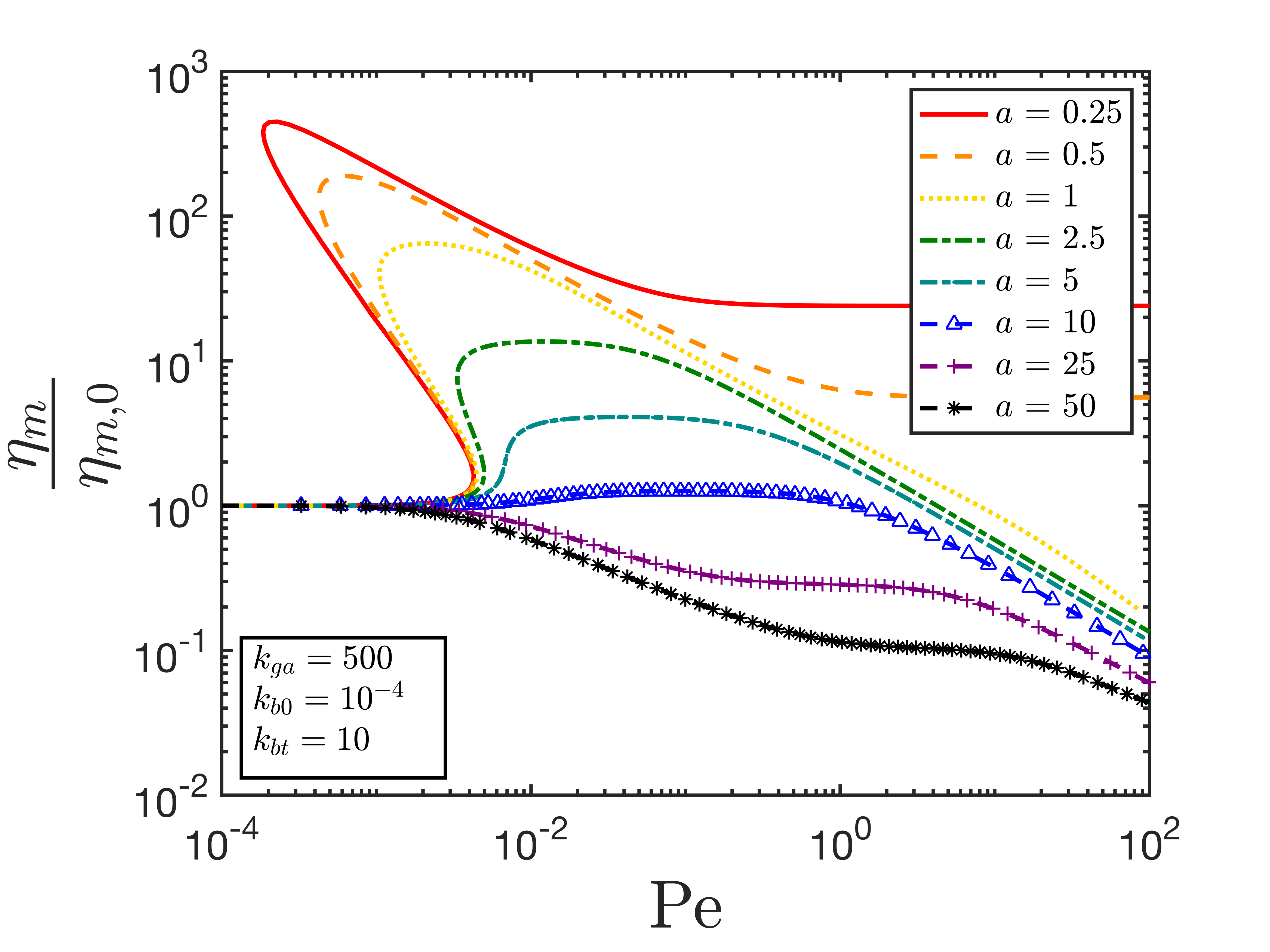}  
        \label{fig:SteadyShear_a_visc}
    \end{subfigure}%
    \begin{subfigure}{.5\textwidth}
        \centering
        \subcaption{}
        \vspace{-3mm}
        \includegraphics[width=.95\linewidth]{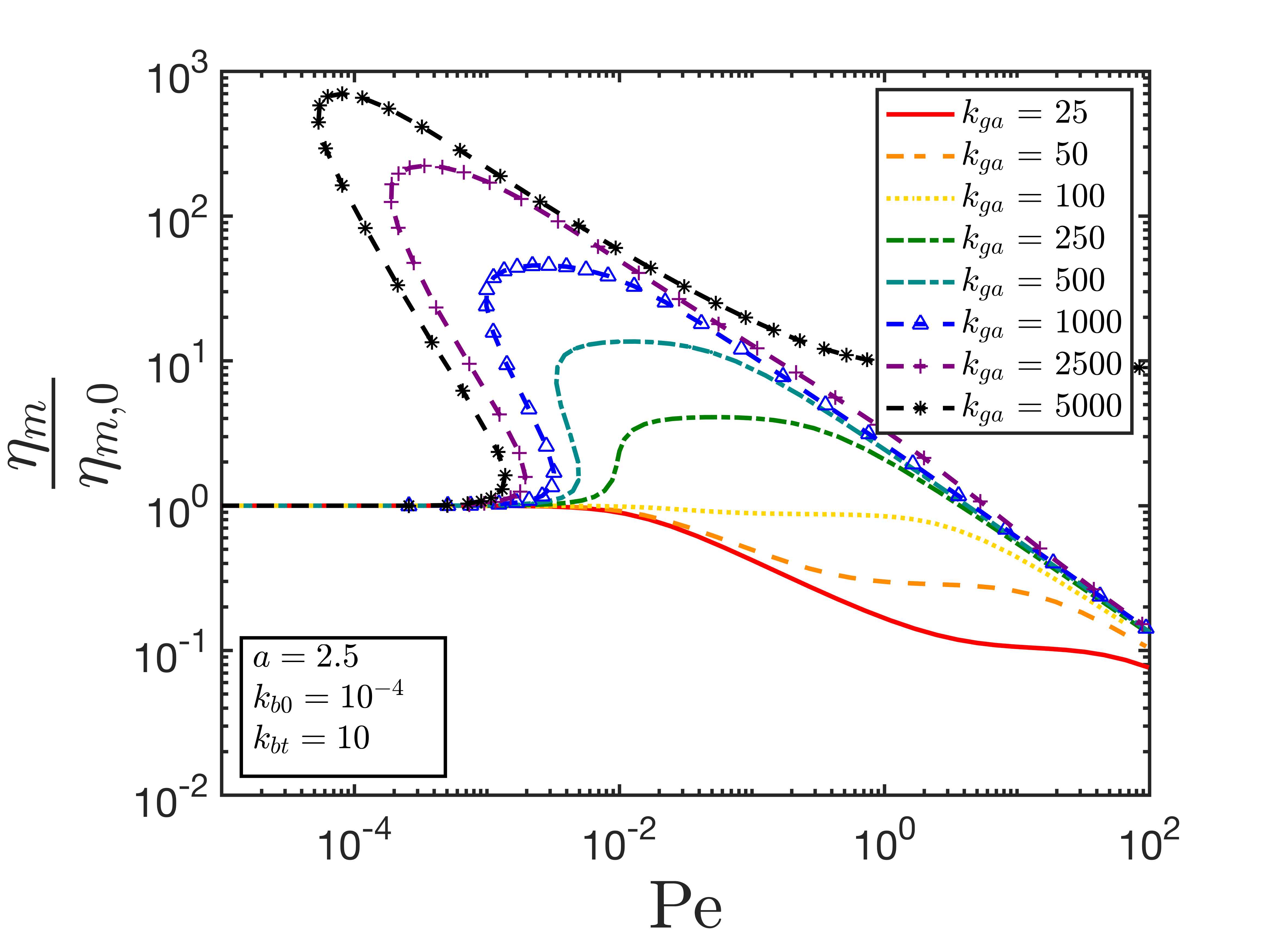}  
        \label{fig:SteadyShear_kga_visc}
    \end{subfigure}
    \begin{subfigure}{.5\textwidth}
        \centering
        \subcaption{}
        \vspace{-3mm}
        \includegraphics[width=.95\linewidth]{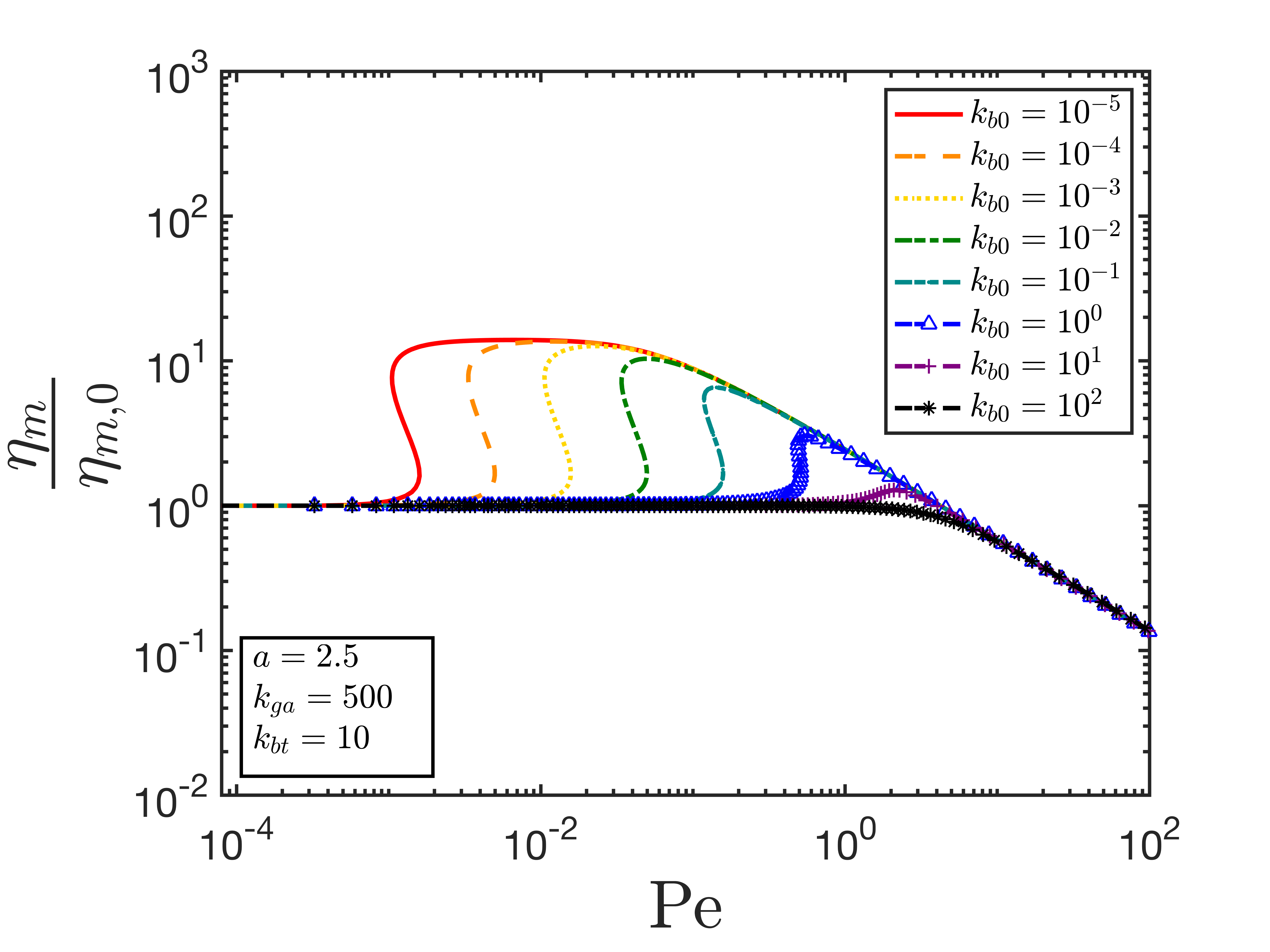}  
        \label{fig:SteadyShear_kb0_visc}
    \end{subfigure}%
    \begin{subfigure}{.5\textwidth}
        \centering
        \subcaption{}
        \vspace{-3mm}
        \includegraphics[width=.95\linewidth]{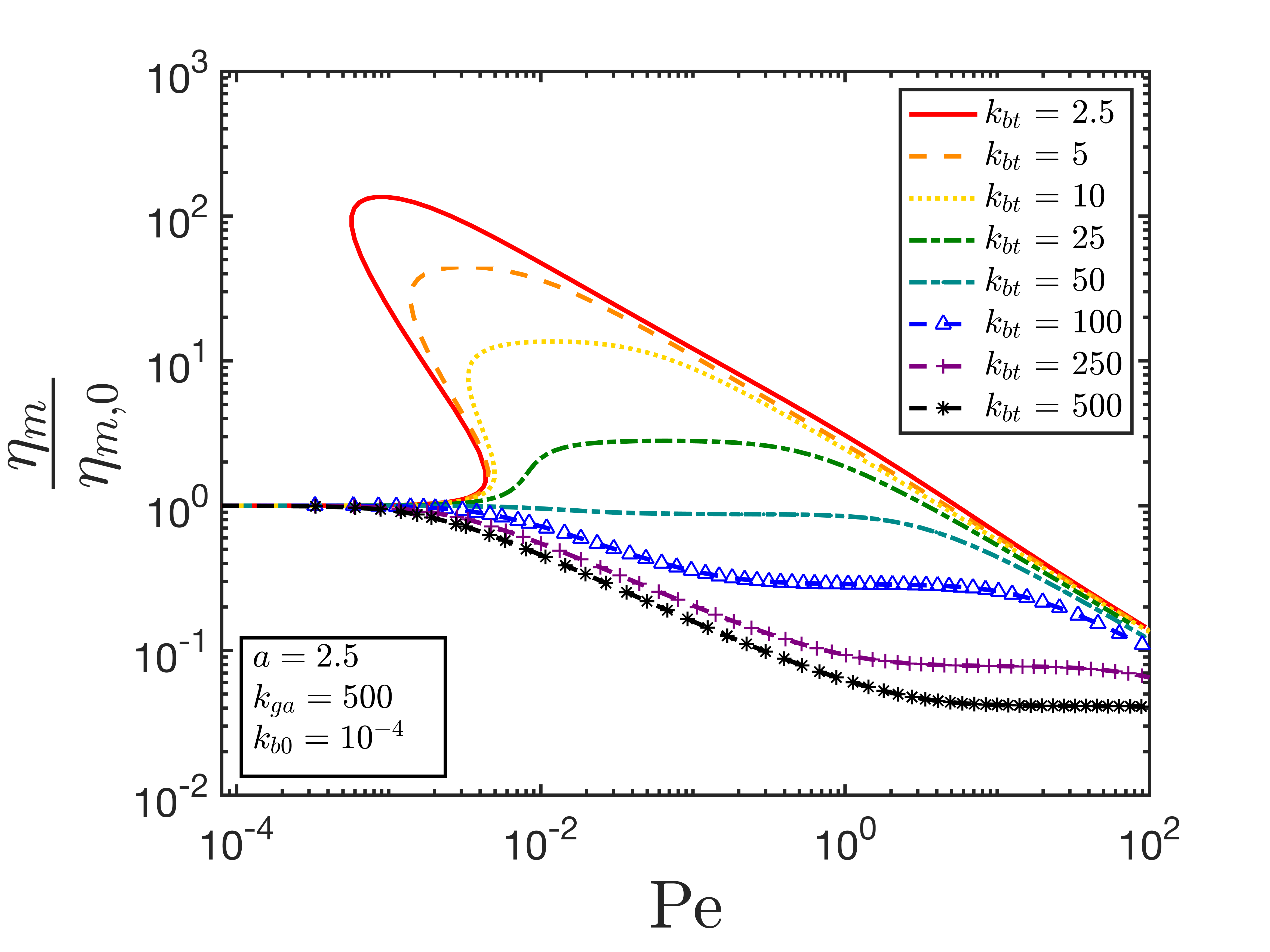}  
        \label{fig:SteadyShear_kbf_visc}
    \end{subfigure}
    \caption{Normalized micellar viscosity vs. shear rate in simple shear flow: \subref{fig:SteadyShear_a_visc} effect of $a$, \subref{fig:SteadyShear_kga_visc} effect of $k_{ga}$, \subref{fig:SteadyShear_kb0_visc} effect of $k_{b0}$, and \subref{fig:SteadyShear_kbf_visc} effect of $k_{bt}$ with $m = 3$ and $D_{r,0} = 1$.}
    \label{fig:SteadyShearVisc}
\end{figure*}

\begin{figure*}
    \begin{subfigure}{.5\textwidth}
        \centering
        \subcaption{}
        \vspace{-3mm}
        \includegraphics[width=.95\linewidth]{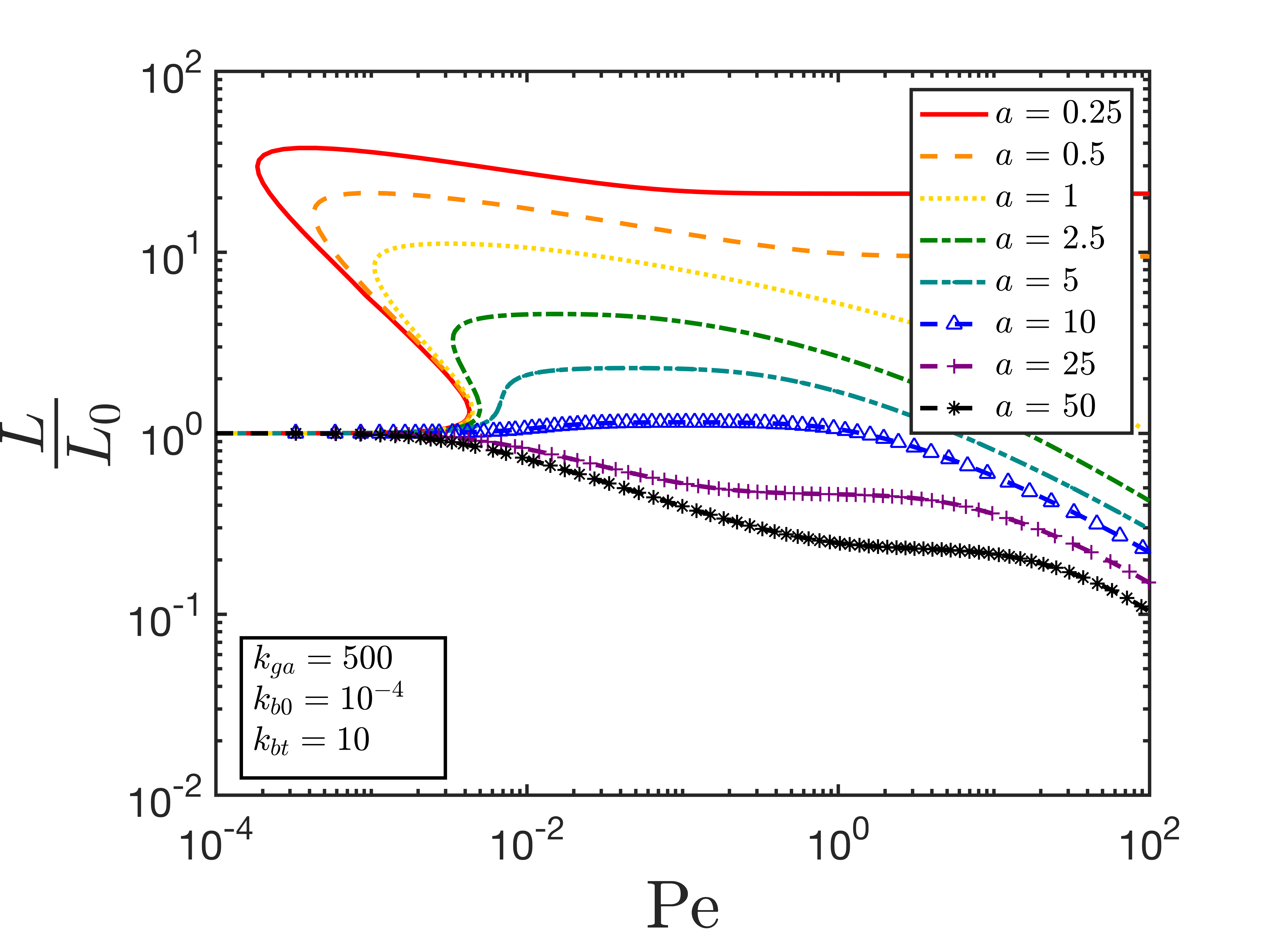}  
        \label{fig:SteadyShear_a_L}
    \end{subfigure}%
    \begin{subfigure}{.5\textwidth}
        \centering
        \subcaption{}
        \vspace{-3mm}
        \includegraphics[width=.95\linewidth]{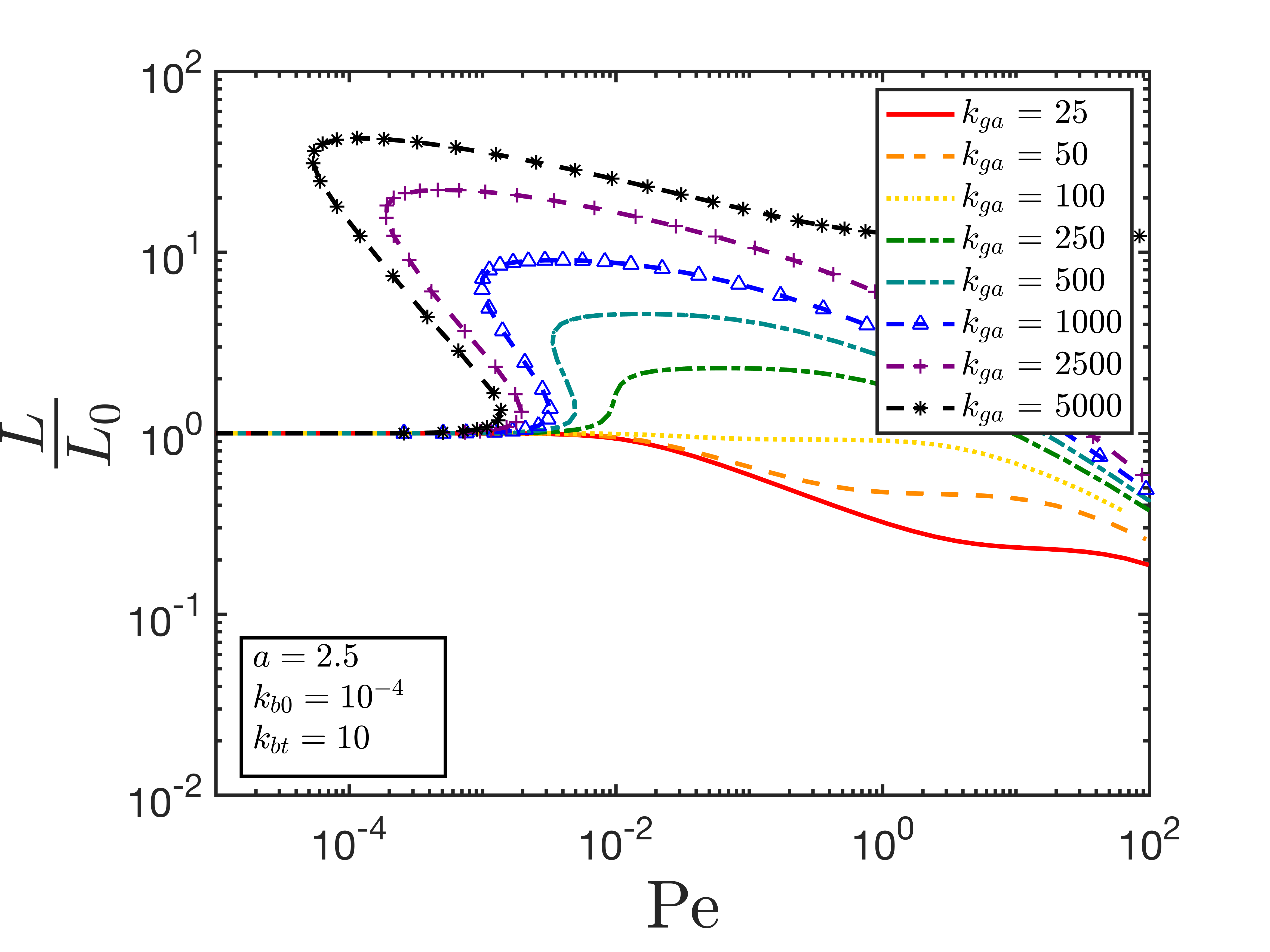}  
        \label{fig:SteadyShear_kga_L}
    \end{subfigure}
    \begin{subfigure}{.5\textwidth}
        \centering
        \subcaption{}
        \vspace{-3mm}
        \includegraphics[width=.95\linewidth]{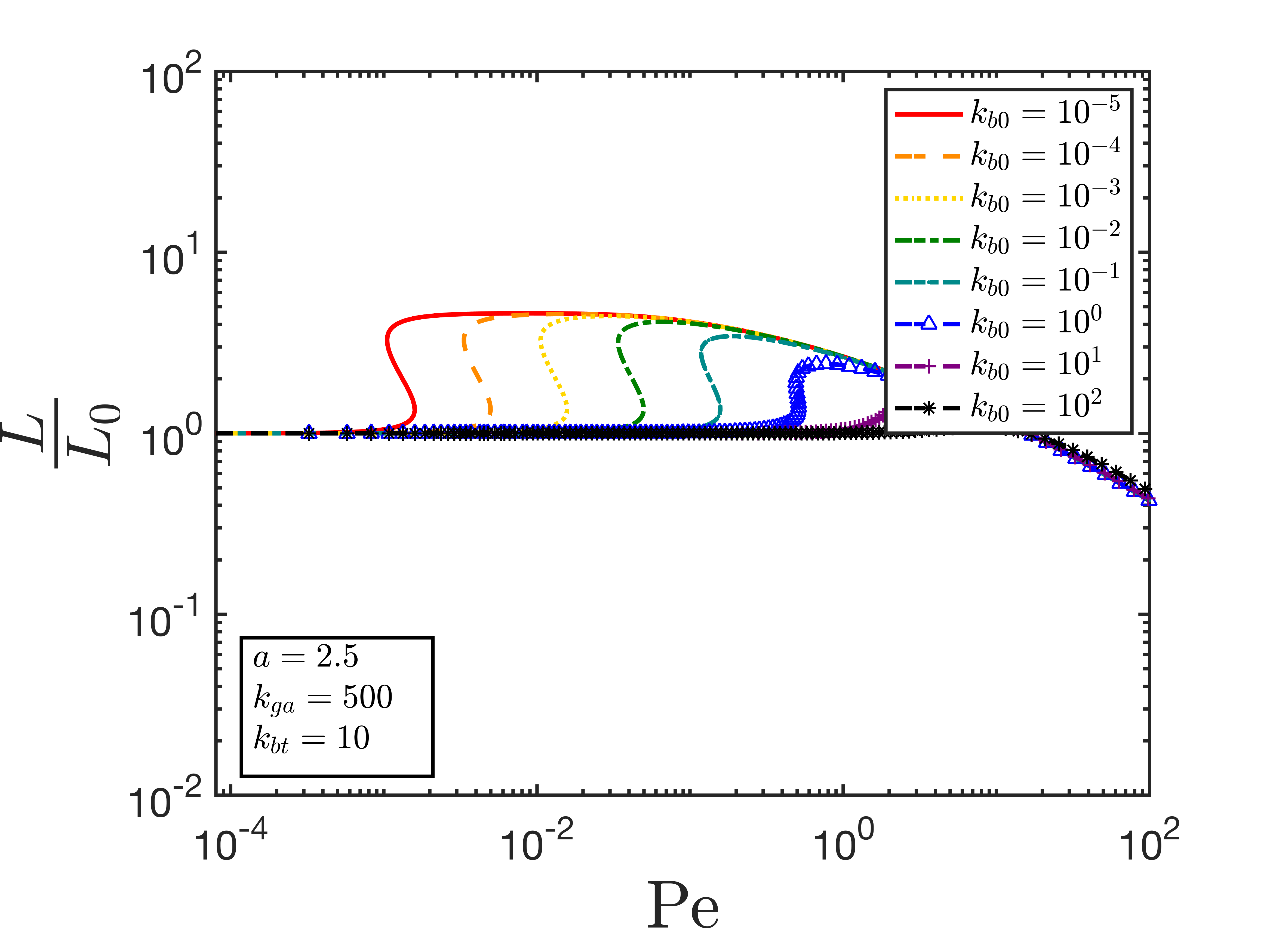}  
        \label{fig:SteadyShear_kb0_L}
    \end{subfigure}%
    \begin{subfigure}{.5\textwidth}
        \centering
        \subcaption{}
        \vspace{-3mm}
        \includegraphics[width=.95\linewidth]{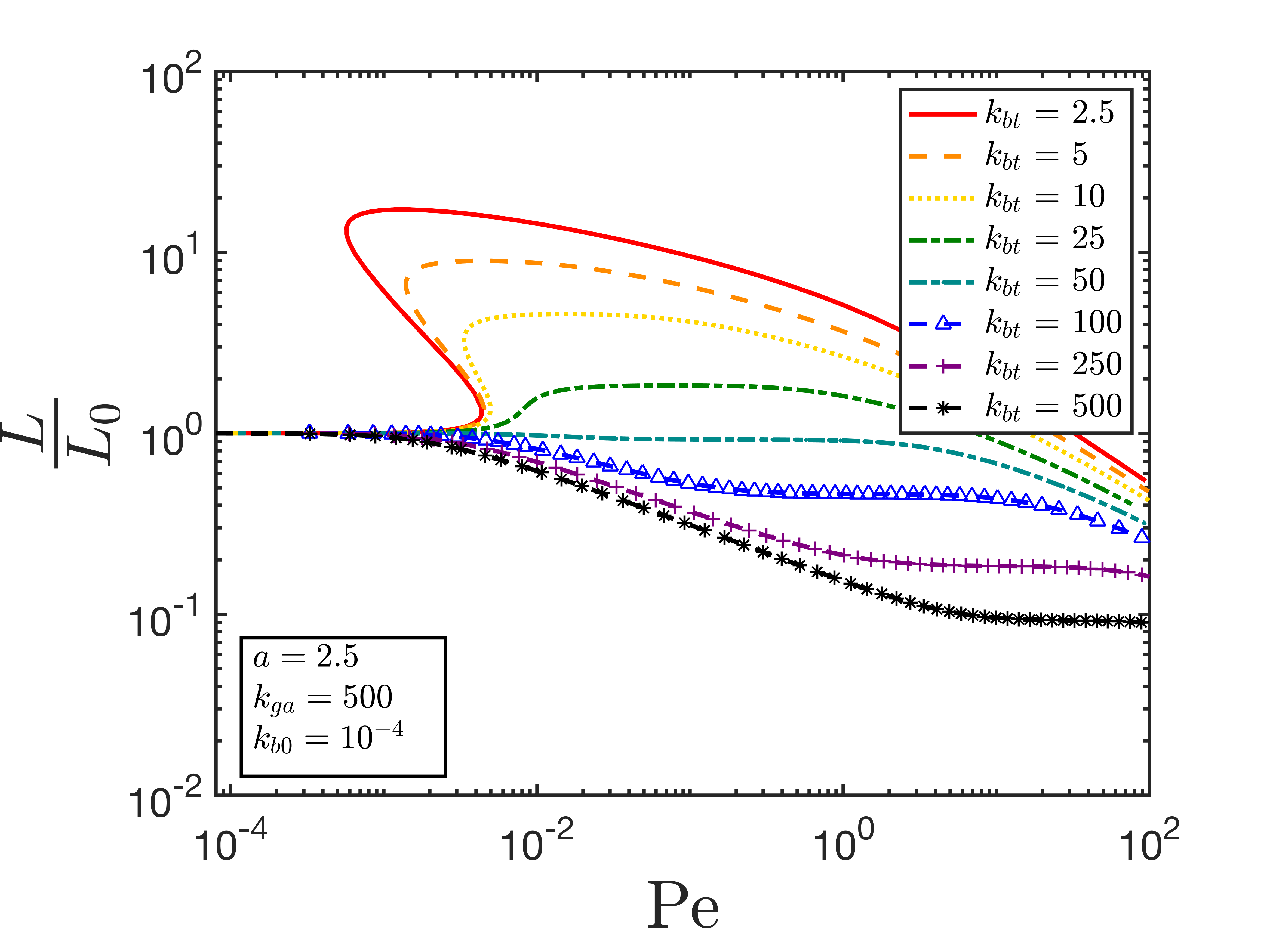}  
        \label{fig:SteadyShear_kbf_L}
    \end{subfigure}
    \caption{Mean micelle length normalized by equilibrium length vs. shear rate in simple shear flow: \subref{fig:SteadyShear_a_L} effect of $a$, \subref{fig:SteadyShear_kga_L} effect of $k_{ga}$, \subref{fig:SteadyShear_kb0_L} effect of $k_{b0}$, and \subref{fig:SteadyShear_kbf_L} effect of $k_{bt}$ with $m = 3$ and $D_{r,0} = 1$.}
    \label{fig:SteadyShearLength}
\end{figure*}

In this section we survey the behavior of the model over a wide range of parameters. Our aim here is not to capture specific experimental observations -- that will be done in the subsequent sections. \RJHrevise{We have combined the discussions of the normalized micelle contribution to the viscosity ($\eta_m/\eta_{m,0}$), \cref{fig:SteadyShearVisc}, and normalized micelle length ($L/L_0$), \cref{fig:SteadyShearLength}, as functions  of \Peclet{} number ($\mathrm{Pe} = \dot{\gamma}/D_{r,0}$) as they tend to show nearly identical parameter dependence. Note that for convenience we have dropped the asterisks from all dimensionless parameters for the remainder of this manuscript. \Cref{fig:SteadyShear_a_visc,fig:SteadyShear_a_L} show the effect of the exponential breakage parameter $a$ with $k_{b0} = 10^{-4}$, $k_{ga} = 500$, and $k_{bt} = 10$; we see that increasing the value of $a$ decreases the magnitude of shear-thickening and micelle elongation and as $a$ is increased beyond some limit prescribed by a combination of the other three parameters (here $a \gtrsim 10$), the suspension transitions to a purely shear-thinning fluid. In this shear-thinning regime ($a = 25, 50$) we see that the suspension undergoes distinctly two thinning regimes separated by a plateau that extends over almost a decade of shear rates. This result reflects the fact that two mechanisms for shear-thinning exist in the model: alignment of the rods with flow and tension-induced rod breakage. \RJHrevise{To the best of our knowledge this plateau behavior that extends into the high-shear rate regime is not typically observed in experiments; however, a similar plateau in the intermediate-shear rate regime has been observed \cite{Wu2019}, though at high enough shear rates this eventually gives way to shear-thinning.}}

\RJHrevise{In regimes where the suspension shear-thickens ($a \lesssim 10$), the viscosity can be seen to thicken to over 100 times its zero-shear value and the micelle length increases by almost 50 times its equilibrium value, consistent with experimental observations of flow-induced structures in dilute wormlike micelle solutions \cite{Lerouge2009,Boltenhagen1997}. This regime contains a significant transition where values of $a$ below some threshold (here $\approx 4$) exhibit reentrant behavior (i.e. multivalued stress-shear rate curve).} In this reentrant region the fluid can take on three steady state stresses for a given shear rate; all three of the stress values are accessible via a strain-controlled experiment, whereas a shear rate-controlled experiment will exhibit hysteresis and jump between the minimum and maximum stresses, leaving the intermediate stress inaccessible. Finally, we see that for small enough values of $a$ ($\lesssim 1$) the viscosity and length plateau at high \Peclet{} numbers. Intuitively we expect high shear rates and stresses to break apart micelles, leading to lengths (and consequently viscosities) that asymptotically approach zero. While this may be nonphysical behavior, it is easily avoided by ensuring $a$ exceeds a necessary threshold.

\Cref{fig:SteadyShear_kga_visc,fig:SteadyShear_kga_L} show the effect of the alignment-induced growth parameter $k_{ga}$ with $a = 2.5$, $k_{b0} = 10^{-4}$, and $k_{bt} = 10$. The effect of increasing $k_{ga}$ is nearly equivalent to decreasing $a$, where larger values of $k_{ga}$ (and smaller values of $a$) induce stronger shear-thickening and micelle elongation. At $k_{ga} = 5000$ we see that the viscosity increases to nearly 1000-times its zero-shear value; we also see for this parameter value that the viscosity and length plateau at high \Peclet{} numbers, which although nonphysical can be avoided by tuning the relationship between $a$ and $k_{ga}$. The effects of the breakage coefficient $k_{bt}$ are shown in \cref{fig:SteadyShear_kbf_visc,fig:SteadyShear_kbf_L} and very nearly mirror the effects of $a$ so we will not discuss them in greater detail.

\begin{figure*}[t]
    \begin{subfigure}{.5\textwidth}
        \centering
        \subcaption{}
        \vspace{-3mm}
        \includegraphics[width=.95\linewidth]{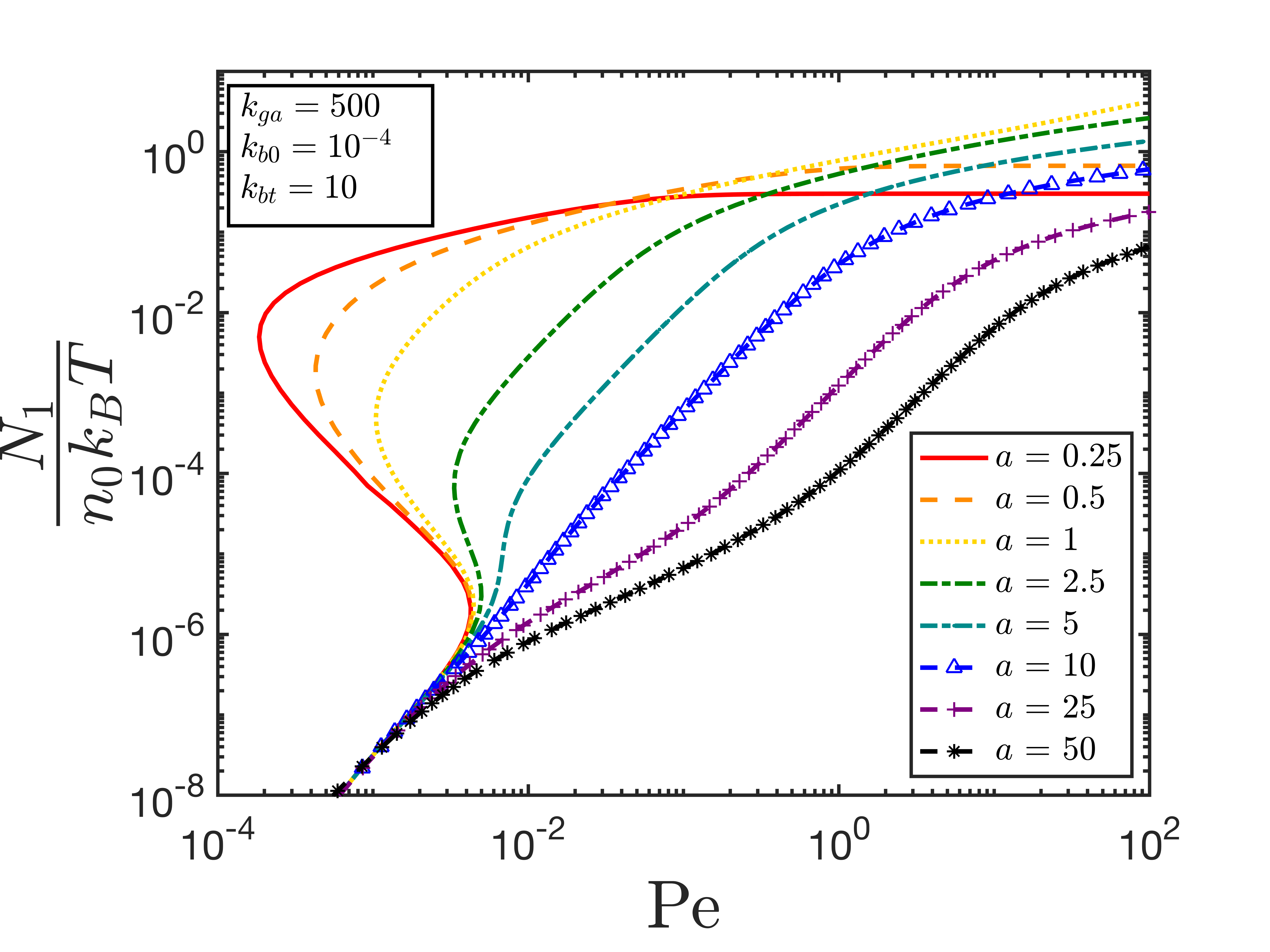}  
        \label{fig:SteadyShear_a_N1}
    \end{subfigure}%
    \begin{subfigure}{.5\textwidth}
        \centering
        \subcaption{}
        \vspace{-3mm}
        \includegraphics[width=.95\linewidth]{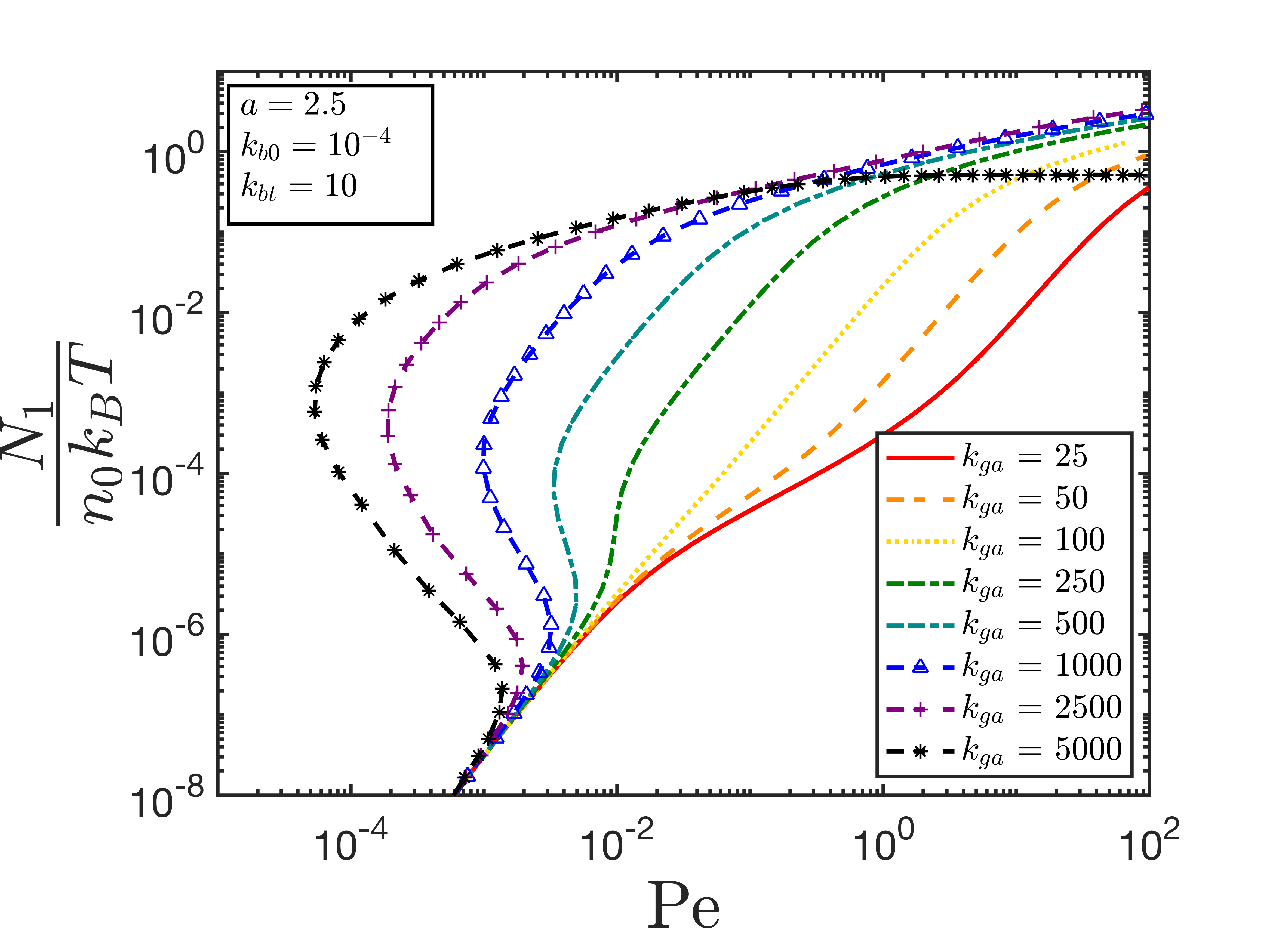}  
        \label{fig:SteadyShear_kga_N1}
    \end{subfigure}
    \begin{subfigure}{.5\textwidth}
        \centering
        \subcaption{}
        \vspace{-3mm}
        \includegraphics[width=.95\linewidth]{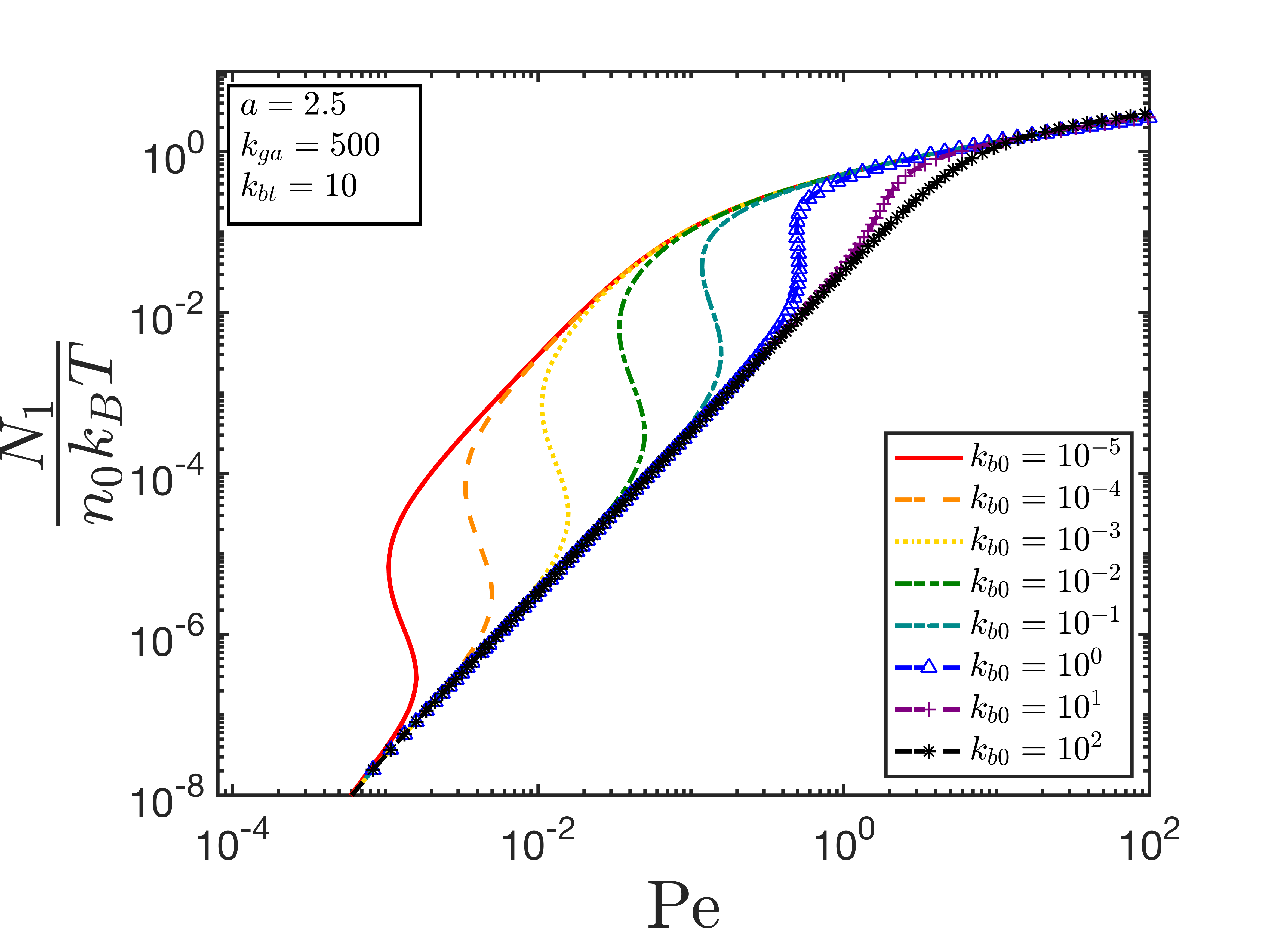}  
        \label{fig:SteadyShear_kb0_N1}
    \end{subfigure}%
    \begin{subfigure}{.5\textwidth}
        \centering
        \subcaption{}
        \vspace{-3mm}
        \includegraphics[width=.95\linewidth]{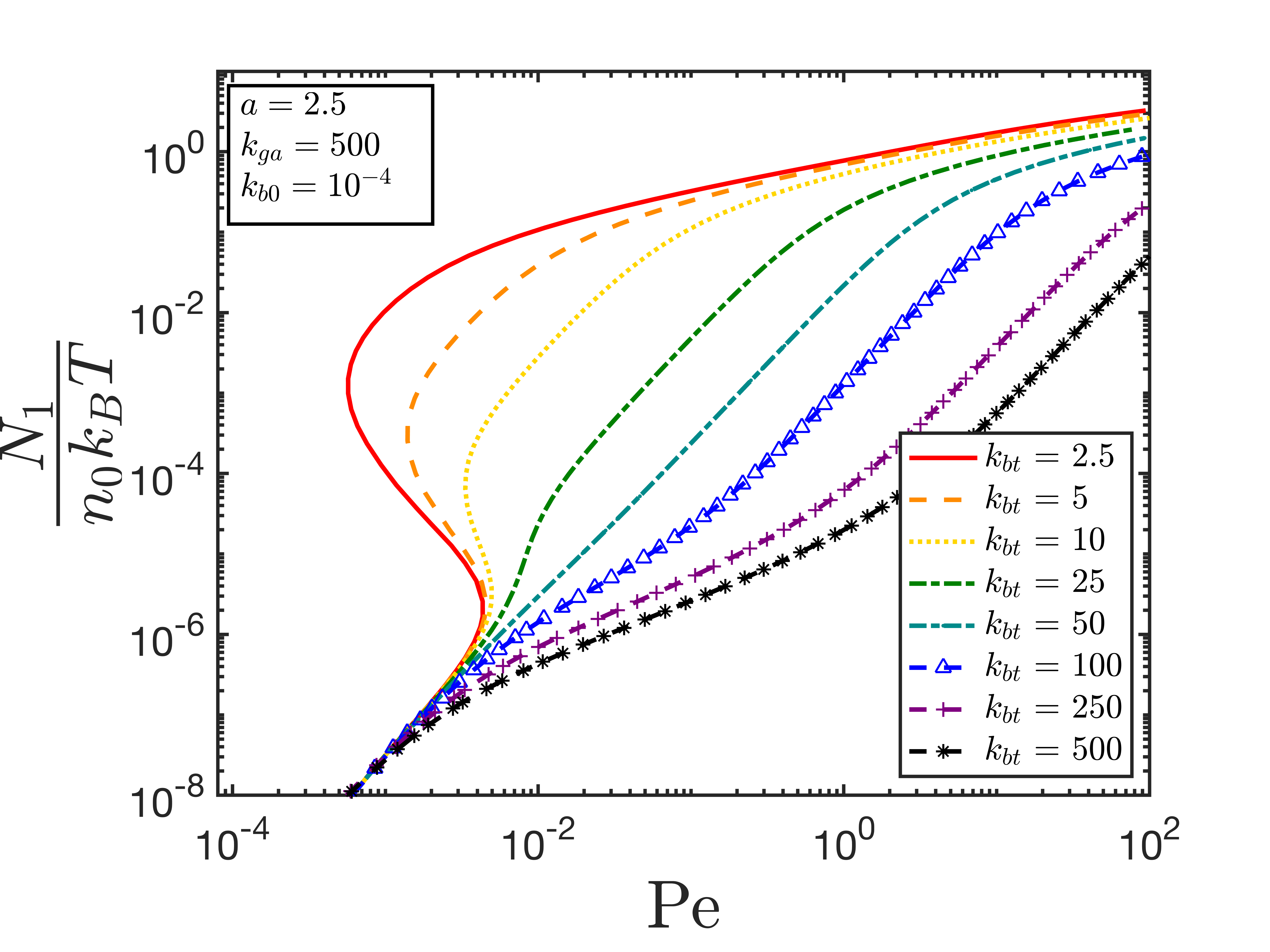}  
        \label{fig:SteadyShear_kbf_N1}
    \end{subfigure}
    \caption{Normalized first normal stress difference vs. shear rate in simple shear flow: \subref{fig:SteadyShear_a_N1} effect of $a$, \subref{fig:SteadyShear_kga_N1} effect of $k_{ga}$, \subref{fig:SteadyShear_kb0_N1} effect of $k_{b0}$, and \subref{fig:SteadyShear_kbf_N1} effect of $k_{bt}$ with $m = 3$ and $D_{r,0} = 1$.}
    \label{fig:SteadyShearN1}
\end{figure*}
The most unique parameter effect can be seen by varying the spontaneous breakage parameter $k_{b0}$ in \cref{fig:SteadyShear_kb0_visc,fig:SteadyShear_kb0_L}. We see that increasing $k_{b0}$ increases the critical shear rate for shear-thickening and elongation to occur, but not necessarily the magnitude of thickening or extent of elongation. Recalling that $k_{b0}$ describes the ratio of relaxation due to breakage to relaxation due to rotational diffusion, this increase in the critical shear rate can be understood by the fact that larger values of $k_{b0}$ correspond to a greater propensity for undergoing a breakage event rather than rotation, and thus a system will prefer to relax by breaking rather than aligning with the flow. An equivalent effect is achieved by decreasing the rotational diffusion constant $D_{r,0}$, which acts to restrict the rods from aligning and in turn restricts micellar growth. The effect of $k_{b0}$ on the magnitude of thickening and elongation is somewhat complicated: at large values of $k_{b0}$, increasing $k_{b0}$ further acts to decrease the magnitude of thickening, while at smaller values ($k_{b0} \lesssim 10^{-3}$) there is no effect on the magnitude of thickening, only the critical shear rate. Notably, different behavior due to variations in $k_{b0}$ can only be seen at low shear rates, while at high shear rates all curves collapse onto identical shear-thinning profiles. This collapse is the result of increased stress that drives the breakdown of elongated micellar structures. 

A recent study by Tamano et al. \cite{Tamano2020} that coupled a fluidity equation to well-known constitutive equations (e.g. Giesekus and FENE-P) to produce both thinning and thickening behaviors (see \cref{sec:Introduction}) identified a nearly identical effect when varying their parameter $R_{bd}$. Similar to $k_{b0}$ in the RRM-R, $R_{bd}$ represents the ratio of the micelle breakdown timescale ($\lambda_{bd}$) to the relaxation time of the fluid ($\lambda$). The fact that these two parameters, which are intended to represent equivalent ratios but exist in two distinct models, yielded extremely similar behavior is quite interesting and worthy of further exploration.

All parameter regimes in \cref{fig:SteadyShearVisc}, excluding curves that show high shear-plateaus, undergo the same power-law thinning $\eta \propto \dot{\gamma}^n$ with $n = -2/3$ at high shear rates. This is in contrast to the high-shear behavior of simple Brownian rods of infinite aspect ratio, which exhibit power-law thinning with $n = -1/3$ \cite{Hinch1972,Ottinger1988}. At finite aspect ratio the viscosity plateaus as $\Pe\rightarrow\infty$. These behaviors are not well-captured using closures, because while the rods spend most of their time aligned with the flow, the high-shear viscosity is dominated by the stress arising during the infrequent flipping of the rods, an effect not accounted for in closures. (The $n=-2/3$ scaling would be correct for the simple rigid rod case if the stress were dominated by the aligned rods.) We are unaware of experimental or modeling results that indicate what the correct scaling behavior should be at very high $\Pe$. We note that we are able to achieve good agreement with the degree of shear thinning found in the experiments shown below.


\Cref{fig:SteadyShearN1} shows the role of the four parameters on the normalized first normal stress difference $N_1/n_0k_BT$, where $N_1 = \tau_{xx}-\tau_{yy}$, of the suspension as a function of \Peclet{} number. \Cref{fig:SteadyShear_a_N1,fig:SteadyShear_kbf_N1} show that increasing $a$ and $k_{bt}$ act to reduce $N_1$ and transition the curve away from reentrant behavior. In \cref{fig:SteadyShear_kga_N1} we have that opposite effect, increasing $k_{ga}$ increases $N_1$ and develops a reentrant profile. \Cref{fig:SteadyShear_kb0_N1} shows that increasing $k_{b0}$ forces the increase of the normal stress difference to higher \Peclet{} numbers, consistent with \cref{fig:SteadyShearVisc}. This plot clearly shows two power-law relationships between the first normal stress and \Peclet{} number, where at low-to-intermediate shear rates $N_1 \propto \mathrm{Pe}^2$ which transitions at high shear rates to $N_1 \propto \mathrm{Pe}^{1/3}$. This is again in contrast to high-shear behavior of non-reactive rods which demonstrate a $N_1 \propto \mathrm{Pe}^{2/3}$ relationship \cite{Ottinger1988}. As discussed above, the origins of this disagreement are as of yet not fully understood but likely stem from either (1) higher-order stress-length coupling or (2) closure approximations. While this power-law relationship is more difficult to observe in \cref{fig:SteadyShear_a_N1,fig:SteadyShear_kga_N1,fig:SteadyShear_kbf_N1}, we can see traces of it for $2.5 \leq a \leq 25$, $100 \leq k_{ga} \leq 1000$, and $10 \leq k_{bt} \leq 100$. This new formulation of the RRM, like the original, is  able to capture non-zero second normal stress differences $N_2$; although we do not show plots here,  $N_2$  follows the same trends as $N_1$ but is negative and roughly two orders of magnitude smaller, which is consistent with other literature reports for viscoelastic fluids \cite{Bird1987}.

\subsection{Steady shear: experimental comparison}

\label{sec:SteadyShearExpr}

\begin{figure}
    \centering
    \begin{subfigure}{.5\textwidth}
        \centering
        \subcaption{}
        \includegraphics[width=.95\linewidth]{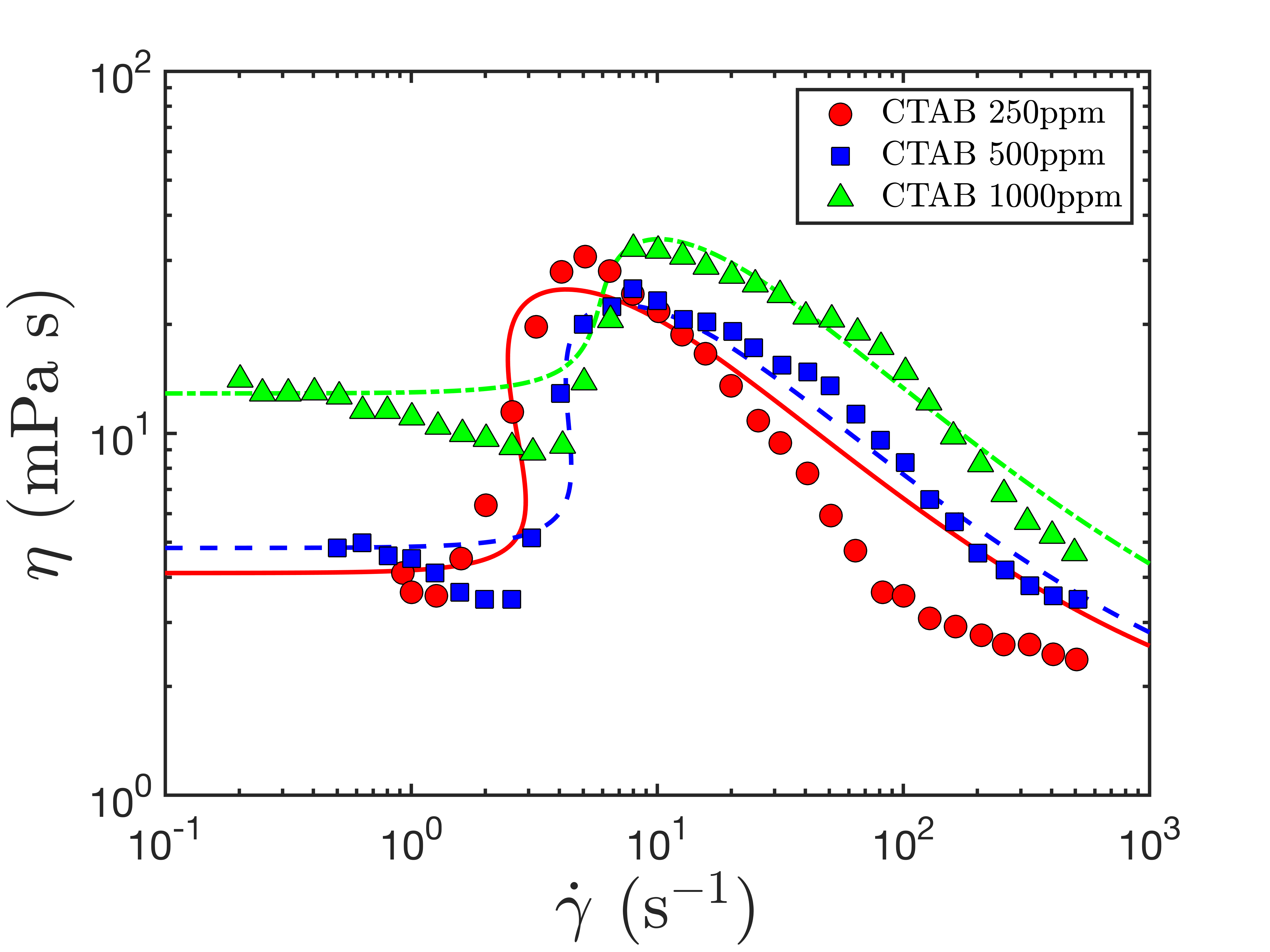}  
        \label{fig:LP_visc}
        \subcaption{}
        \includegraphics[width=.95\linewidth]{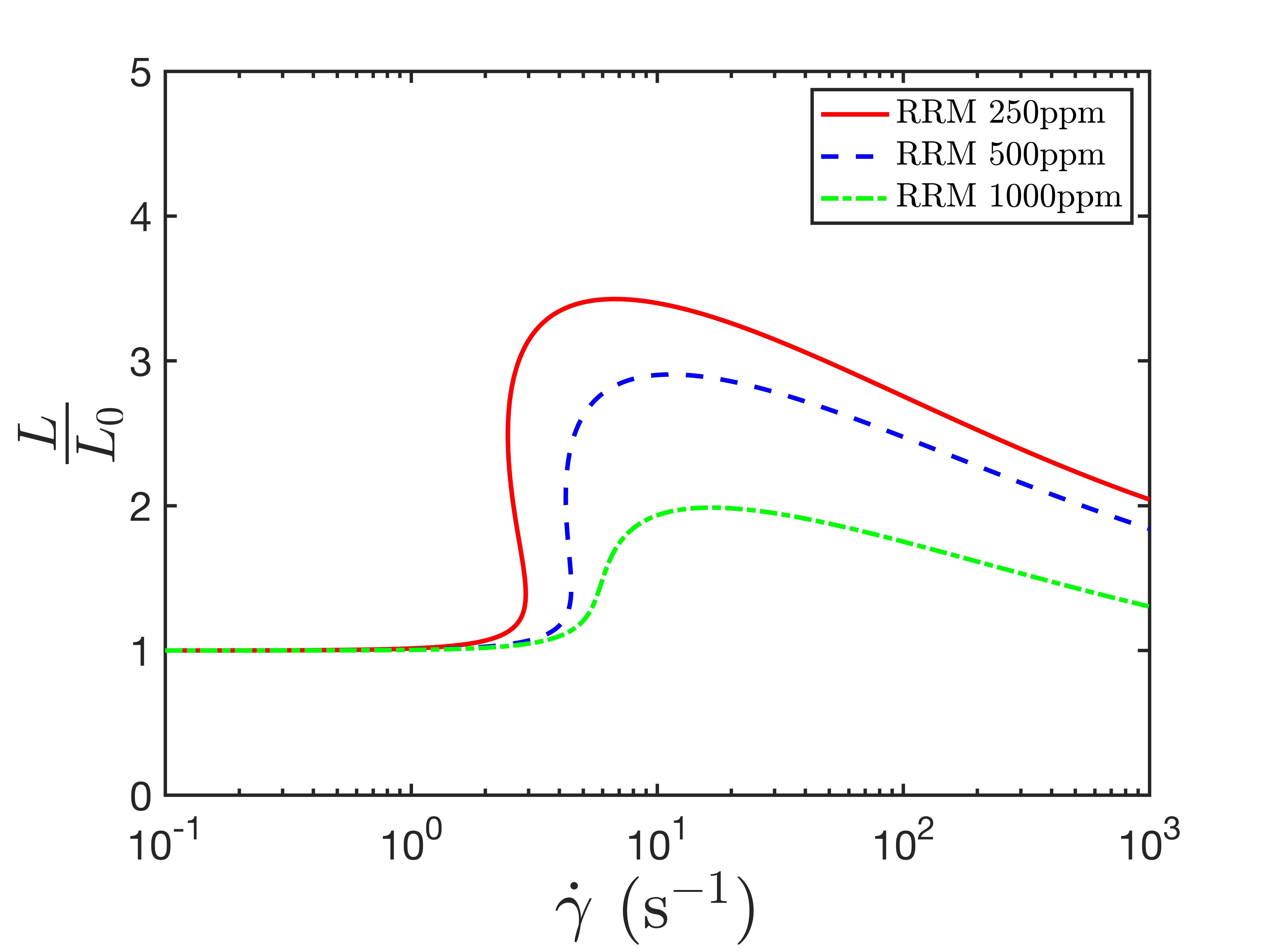}  
        \label{fig:LP_L}
    \end{subfigure}
    \vspace{-3mm}
    \caption{Fits (lines) to experimental data of \subref{fig:LP_visc} shear viscosity vs. shear rate and \subref{fig:LP_L} mean micelle length vs. shear rate. Experimental data corresponds to solutions of 250 ppm, 500 ppm, and 1000 ppm CTAB/NaSal in water obtained by Liu and Pine \cite{Liu1996}. Corresponding parameter values are shown in \cref{table:LP}.}
    \label{fig:LP}
\end{figure}

\begin{figure}
    \centering
    \begin{subfigure}{.5\textwidth}
        \centering
        \subcaption{}
        \includegraphics[width=.95\linewidth]{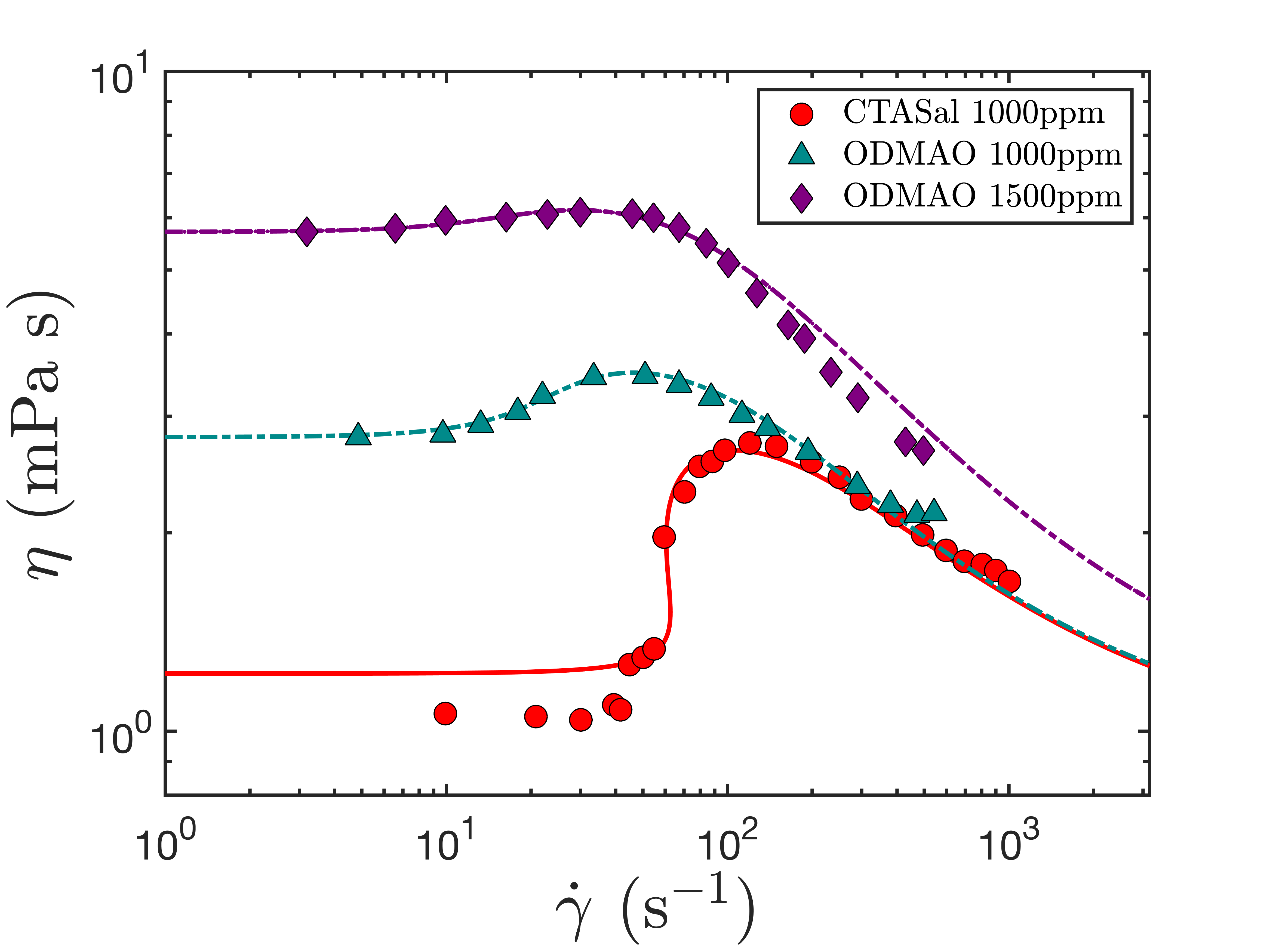}  
        \label{fig:OHTM_visc}
        \subcaption{}
        \includegraphics[width=.95\linewidth]{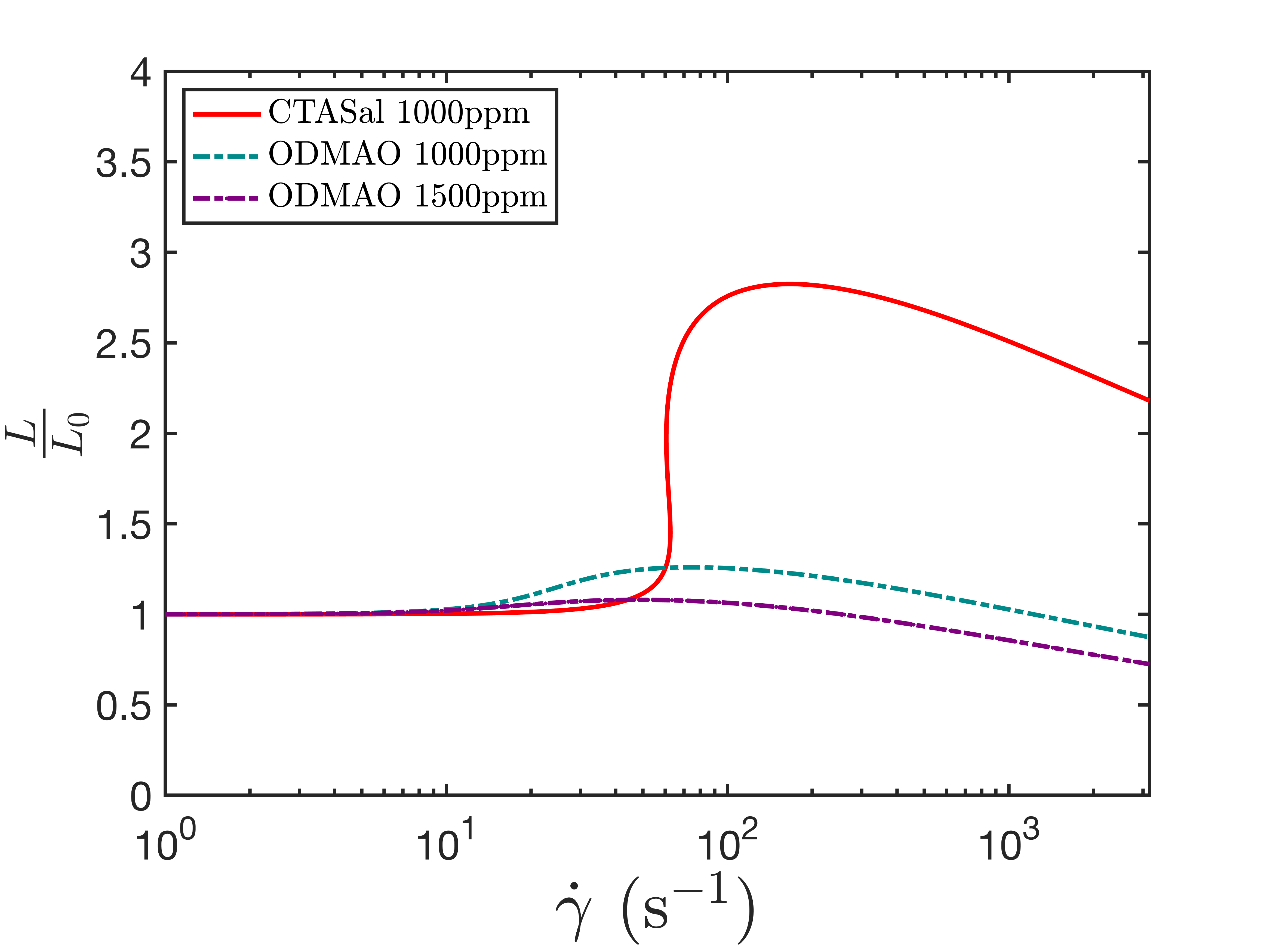}  
        \label{fig:OHTM_L}
    \end{subfigure}
    \vspace{-3mm}
    \caption{Fits (lines) to experimental data of \subref{fig:OHTM_visc} shear viscosity vs. shear rate and \subref{fig:OHTM_L} mean micelle length vs. shear rate. Experimental data corresponds to solutions of (red) 1000 ppm CTASal in water obtained by Ohlendorf et al. \cite{Ohlendorf1986}, and (purple) 1500 ppm and (blue) 1000 ppm ODMAO in water obtained by Tamano et al. \cite{Tamano2017}. Corresponding parameter values are shown in \cref{table:OHTM}.}
    \label{fig:OHTM}
\end{figure}

In this section we turn to comparisons between our model and literature results for steady shear flow.  Model parameters were obtained \RJHrevise{by generating roughly 7$^4$ steady state flow curves (seven different values of each of the four parameters) and plotting the experimental data of interest on each of these curves; we inspected each curve to obtain an approximate `best fit' and then fine-tuned this set by repeating this process with a range of parameters within a few percent of the `best fit' values. We then took the new `best fits' and performed small tweaks to obtain our final fit parameters. We should also note that determination of ideal fitting parameters for the RRM-R is a nonlinear optimization problem and we therefore cannot guarantee the existence of a unique global minimum (i.e. a unique parameter set).} Although we do emphasize that the highly idealized nature of our model means that we should not attach too much physical significance to the values of model parameters, we do see consistency that at least suggests parameters are relatively constant for a given solution composition (e.g. CTAB/NaSal), with the greatest deviations occurring in the most concentrated (semi-dilute) solutions. Note that in \cref{fig:LP_visc,fig:OHTM_visc,fig:BMP_visc} we are showing the \textit{total} solution viscosity ($\eta = \eta_s + \eta_m$) where the solvent viscosity is taken to be that of water. 

The first data set we consider is for a CTAB/NaSal solution in water, from Liu and Pine \cite{Liu1996}. \Cref{fig:LP_visc} shows fits of our model to steady shear viscosity as a function of shear rate for increasing concentrations. Model parameters are shown in \cref{table:LP}. The measurements were made using a double-wall Couette rheometer with an inner cylinder diameter of $25$ mm and gap width of $1$ mm. We find strong agreement between fits and experimental data, particularly in capturing the onset of shear-thickening and transition to shear-thinning. We are notably unable to capture the weak initial shear-thinning at low shear rates that occurs in semi-dilute surfactant solutions ($\geq 500$ ppm); shear-thinning in this regime is attributed to disruption of micellar networks that can form at equilibrium, a phenomenon that is not considered in our treatment.

\Cref{fig:LP_L} shows model predictions for the normalized mean micelle length as a function of shear rate. We find that our model predicts that the mean micelle length increases by about two times in the most concentrated solution and almost four times in the most dilute. While length was not measured in the study by Liu and Pine, these length changes are consistent with other experimental results \cite{Wunderlich1987}.

\begin{table}
    \centering
    \caption{Dimensional RRM-R parameters for experimental data (\cref{fig:LP}) of CTAB/NaSal solutions \cite{Liu1996}.}
    \vspace{-3mm}
    \begin{tabularx}{0.95\textwidth} { 
                >{\raggedright\arraybackslash}X 
                >{\raggedright\arraybackslash}X 
                >{\raggedright\arraybackslash}X
                >{\raggedright\arraybackslash}X}
     \hline
     Composition & CTASal & CTASal & CTASal \\
     $c$ [ppm] & 250 & 500 & 1000 \\ 
     \hline
     $k_{b0}$ [s$^{-1}$] & $1.4 \times 10^{-2}$ & $2.8 \times 10^{-2}$ & $6.3 \times 10^{-2}$ \\ 
     $k_{ga}$ [$\mu\mathrm{m}^{7}$] & $6.1 \times 10^{-6}$ & $3.9 \times 10^{-6}$ & $2.0 \times 10^{-6}$ \\ 
     $k_{bt}$ [m s$^{-1}$] & $1.1 \times 10^{-4}$ & $1.2 \times 10^{-4}$ & $8.1 \times 10^{-5}$ \\ 
     $a$ [nm] & $1.3$ & $1.6$ & $2.2$ \\ 
     $D_{r,0}$ [s$^{-1}$] & 55 & 55 & 25 \\ 
     \hline
    \end{tabularx}
    \label{table:LP}
\end{table}

We also make comparisons to data from \cite{Ohlendorf1986} for both cationic and non-ionic wormlike micelle solutions.
\Cref{fig:OHTM} shows fits to experimental steady shear viscosity as a function of shear rate
data, as well as micelle length predictions. Red data corresponds to a solution of 1000 ppm cationic CTASal in water, measured in a low-shear, circular Couette viscometer at 20 $^{\circ}$C with a gap width of 0.320 mm \cite{Ohlendorf1986}. Blue and purple data correspond to solutions of 1000 ppm and 1500 ppm non-ionic ODMAO in water, respectively, measured with a capillary viscometer with inner diameter 5.045 mm \cite{Tamano2017}. Our model is able to capture the strong shear-thickening and -thinning regimes, particularly that of the cationic solution. Although we see some difficulty in capturing the zero-shear viscosity of the cationic solution (red), this difference less than 1 mPa s and is thus relatively insignificant. We note that the work by Ohlendorf and coworkers \cite{Ohlendorf1986} shows markedly different behaviors for a given wormlike micelle solution by simply changing the gap width of the circular Couette device, indicating the possible presence of instabilities such as vorticity banding.

It is worth drawing attention to model predictions for the rotational diffusion coefficient. Looking at \cref{table:LP,table:OHTM}, we see that predictions for $D_{r,0}$ range from $\sim 10 \, \text{s}^{-1}$ to $\sim 1000 \, \text{s}^{-1}$, whereas for a typical micelle length of a few hundred angstroms \cref{eqn:DiffusionCoeff} predicts values of $D_{r,0}$ on the order of $10^5\,\text{s}^{-1}$. Further, for rigid rods we can relate the relaxation time of the bulk fluid to the rotational diffusivity by $\lambda = 1/6D_{r,0}$, where for wormlike micelle solutions $\lambda$ is typically on the order of $1-10^1 \, \text{s}$; we can then see that our predictions of $D_{r,0}$ yield relaxation times that fall significantly below experimental reports. This discrepancy between experimental measurements of WLM solution relaxation times and theoretical predictions using \cref{eqn:DiffusionCoeff} and based on the framework proposed by Cates and Turner -- shear-thickening occurs when the shear is strong enough to overcome rotational diffusion -- is well-established in literature \cite{Hofmann1991,Oda2000}. 
Many studies have attempted to address this inconsistency; notably, Barentin and Liu \cite{Barentin2001} proposed a mechanism in which micelles form networks of bundles with some success. In our model, there is strong reason to believe that the disparity between predictions and experimental measurements arises from the neglect of charge and hydrodynamic interactions, as well as certain collision mechanisms. In particular, the assumption that collisions are primarily end-to-end in nature ignores the possibility of `phantom crossings' that have been observed in dissipative particle dynamics studies \cite{Yamamoto2005}.

\begin{table*}
    \centering
    \caption{Dimensional RRM-R parameters for experimental data (\cref{fig:OHTM}) of CTASal and ODMAO solutions obtained by \cite{Ohlendorf1986} and \cite{Tamano2017}, respectively.}
    \vspace{-3mm}
    \begin{tabularx}{0.9\textwidth} { 
        >{\raggedright\arraybackslash}X 
        >{\raggedright\arraybackslash}X 
        >{\raggedright\arraybackslash}X
        >{\raggedright\arraybackslash}X
        >{\raggedright\arraybackslash}X }
    \hline
    Composition & CTASal & ODMAO & ODMAO \\
    $c$ [ppm] & 1000 & 1000 & 1500 \\ 
    \hline
    $k_{b0}$ [s$^{-1}$] & $1.2$ & $2.5 \times 10^{-1}$ & $1.5 \times 10^{-1}$ \\ 
    $k_{ga}$ [$\mu\mathrm{m}^{7}$] & $8.6 \times 10^{-6}$ & $2.2 \times 10^{-5}$ & $7.8 \times 10^{-6}$ \\ 
    $k_{bt}$ [m s$^{-1}$] & $2.0 \times 10^{-3}$ & $1.1 \times 10^{-4}$ & $6.8 \times 10^{-5}$ \\ 
    $a$ [nm] & $3.1$ & $2.5$ & $3.0$\\ 
    $D_{r,0}$ [s$^{-1}$] & 800 & 50 & 30 \\ 
    \hline
    \end{tabularx}
    \label{table:OHTM}
\end{table*}

\subsection{Startup of steady shear}
\label{sec:TransientShear}
One of the primary pitfalls of the previous RRM formulation is its difficulty in capturing the \emph{transient} dynamics of dilute surfactant solutions. Accurately describing transient dynamics is crucial for studying complex and turbulent flows. Additionally, it is important to note that properly fitting experimental data requires both steady \textit{and} transient data. The reason for this requirement is that at steady state, \cref{eqn:LengthDimensionless} can be multiplied by any positive scalar value and still produce the same steady shear viscosity vs. shear rate curve. However, multiplying by a scalar does alter transient data, particularly the induction times ($t_\text{ind}$) and overshoots seen in startup of steady shear flows. Thus, if we wish to completely describe dilute WLM solutions, we must be able to accurate predict both steady and transient data in tandem.

We assess the capabilities of the model at capturing transient dynamics by evaluating its behavior in startup of steady shear flow. An initially isotropic fluid at rest is subjected starting at $t = 0$ to a constant shear rate $\dot{\gamma}$. We use an explicit, fourth-order Runge-Kutta scheme implemented in Matlab (\texttt{ode45}) to time-step \cref{eqn:ShearOrientation,eqn:ShearStress,eqn:ShearLengthDimless,eqn:ShearShat}. In evaluating the success of the RRM-R at predicting transient dynamics, we are looking particularly at induction times, the time required for stress growth to occur, as well as stress overshoot, where the stress (and viscosity) is seen to overshoot its steady state value.

\begin{figure*}
    \begin{subfigure}{.5\textwidth}
        \centering
        \subcaption{}
        \vspace{-3mm}
        \includegraphics[width=1\linewidth]{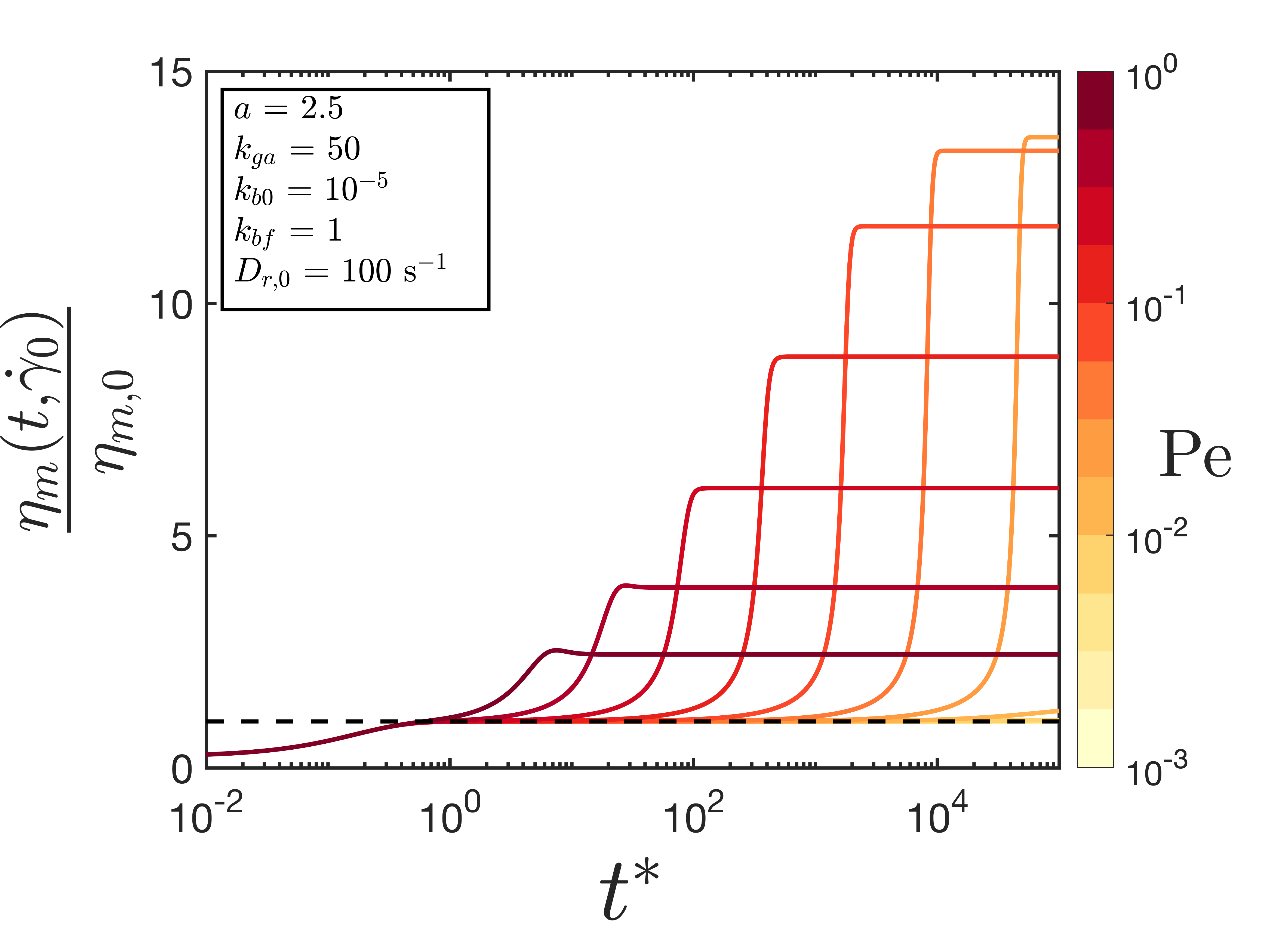}  
    	\label{fig:TransientShearVisc1}
    \end{subfigure}
    \vspace{-10mm}
    \begin{subfigure}{.5\textwidth}
        \subcaption{}
        \vspace{-3mm}
        \includegraphics[width=\linewidth]{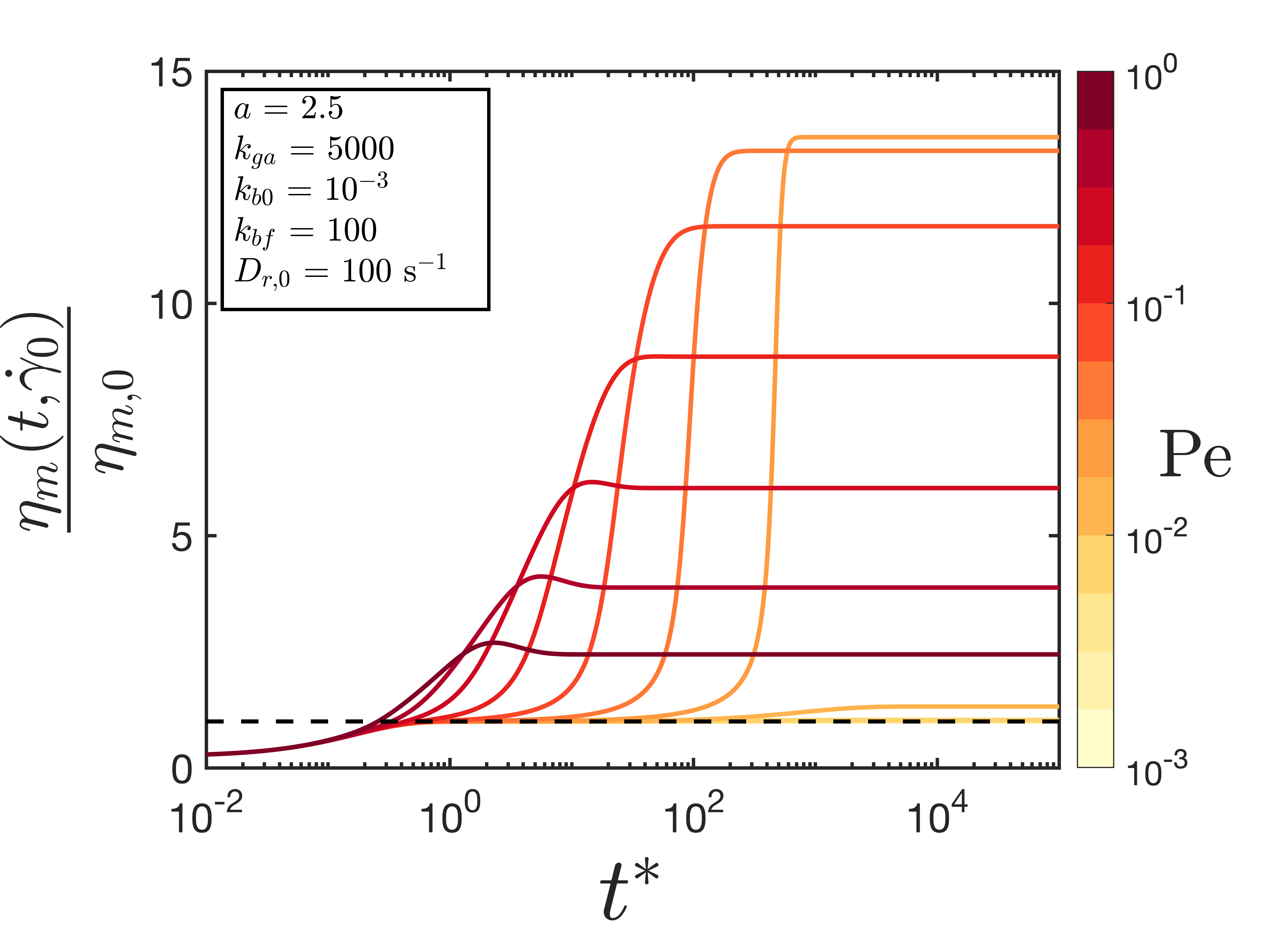}  
    	\label{fig:TransientShearVisc2}
    \end{subfigure}
    
    \begin{subfigure}{.5\textwidth}
        \centering
        \subcaption{}
        \vspace{-3mm}
        \includegraphics[width=1\linewidth]{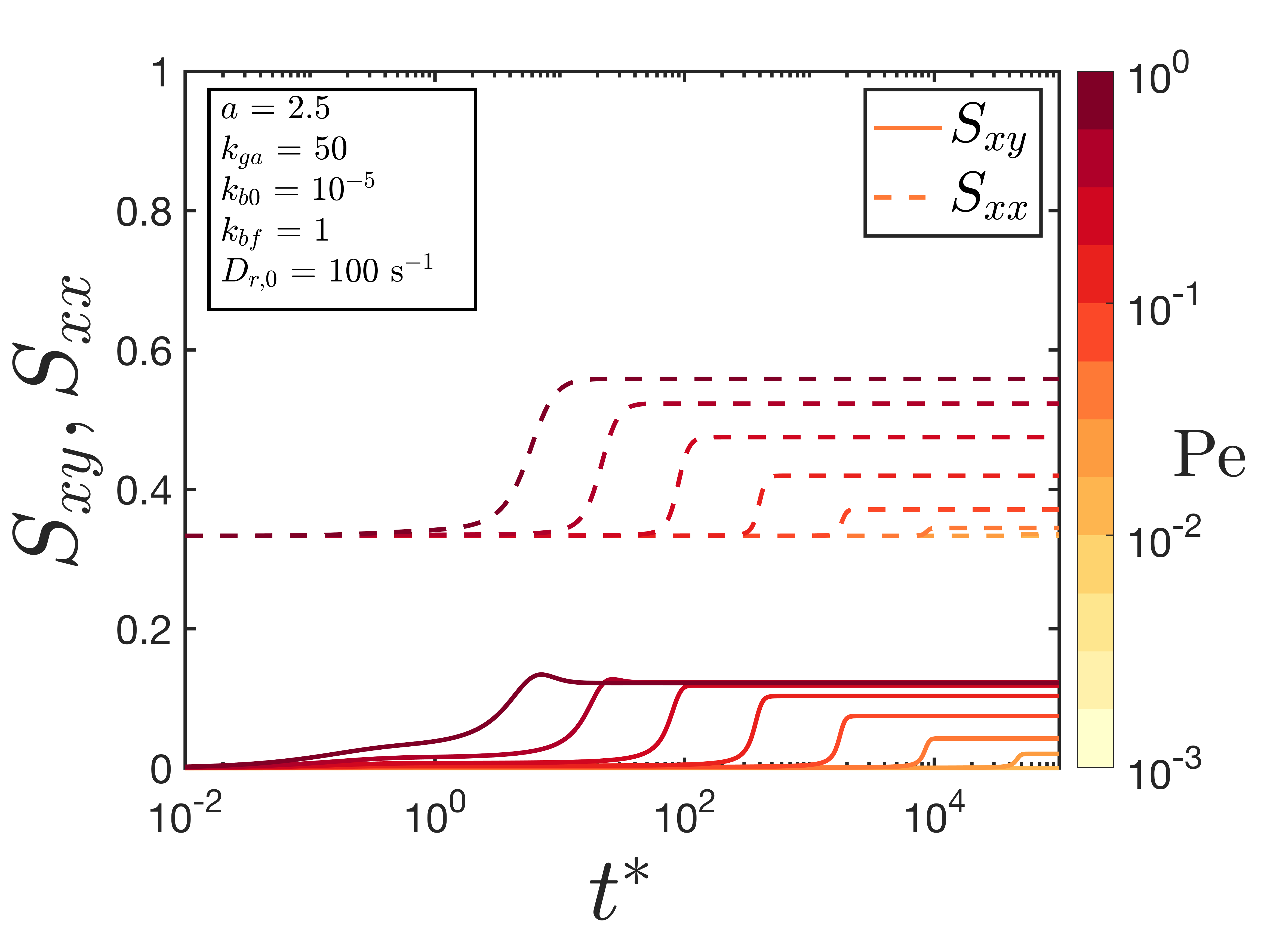} 
        \label{fig:TransientShearOrient1}
    \end{subfigure}
    \vspace{-10mm}
    \begin{subfigure}{.5\textwidth}
    	\centering
        \subcaption{}
        \vspace{-3mm}
        \includegraphics[width=\linewidth]{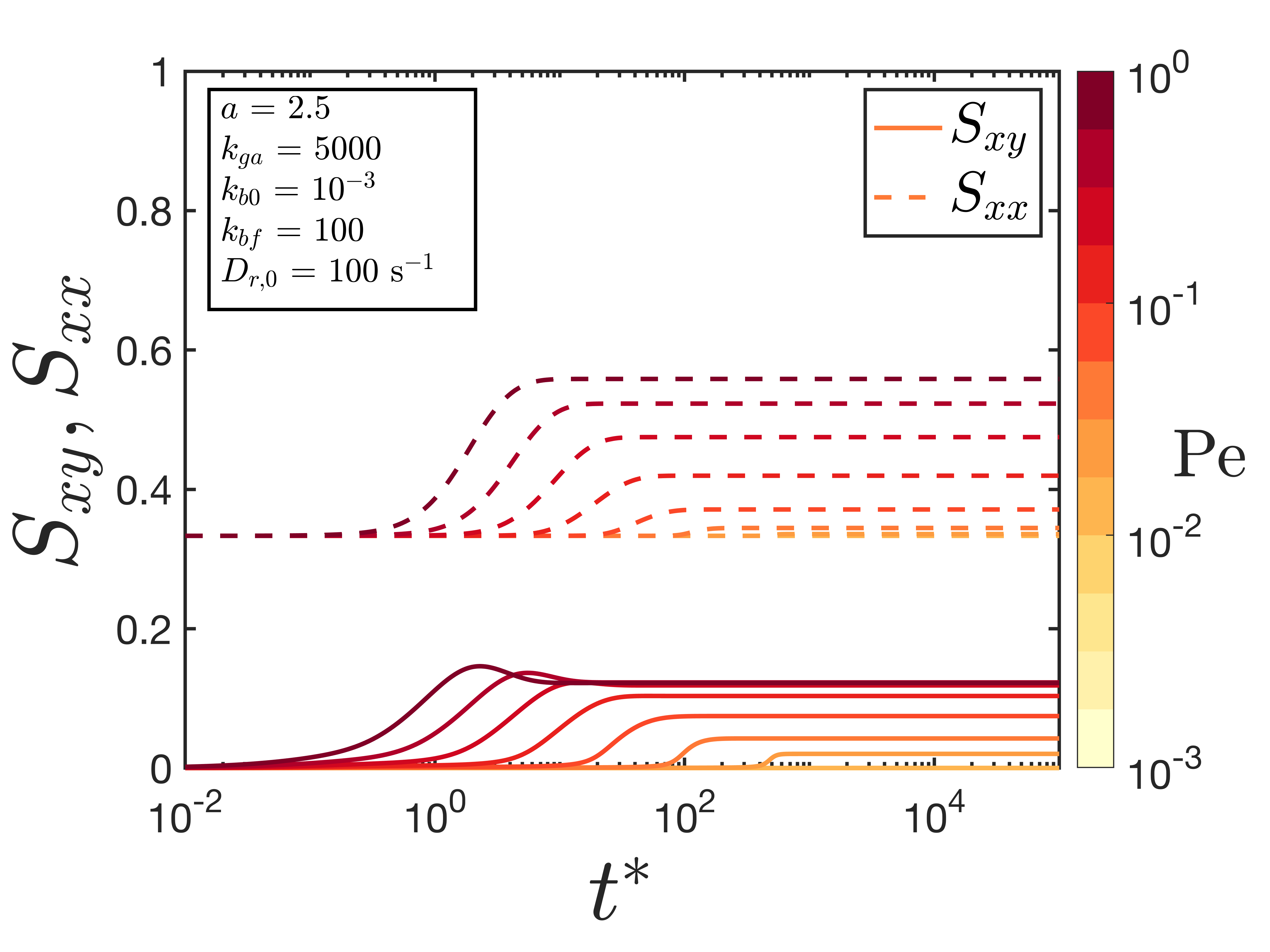}  
        \label{fig:TransientShearOrient2}
    \end{subfigure}
    \begin{subfigure}{.5\textwidth}
        \centering
        \subcaption{}
        \vspace{-3mm}
        \includegraphics[width=1\linewidth]{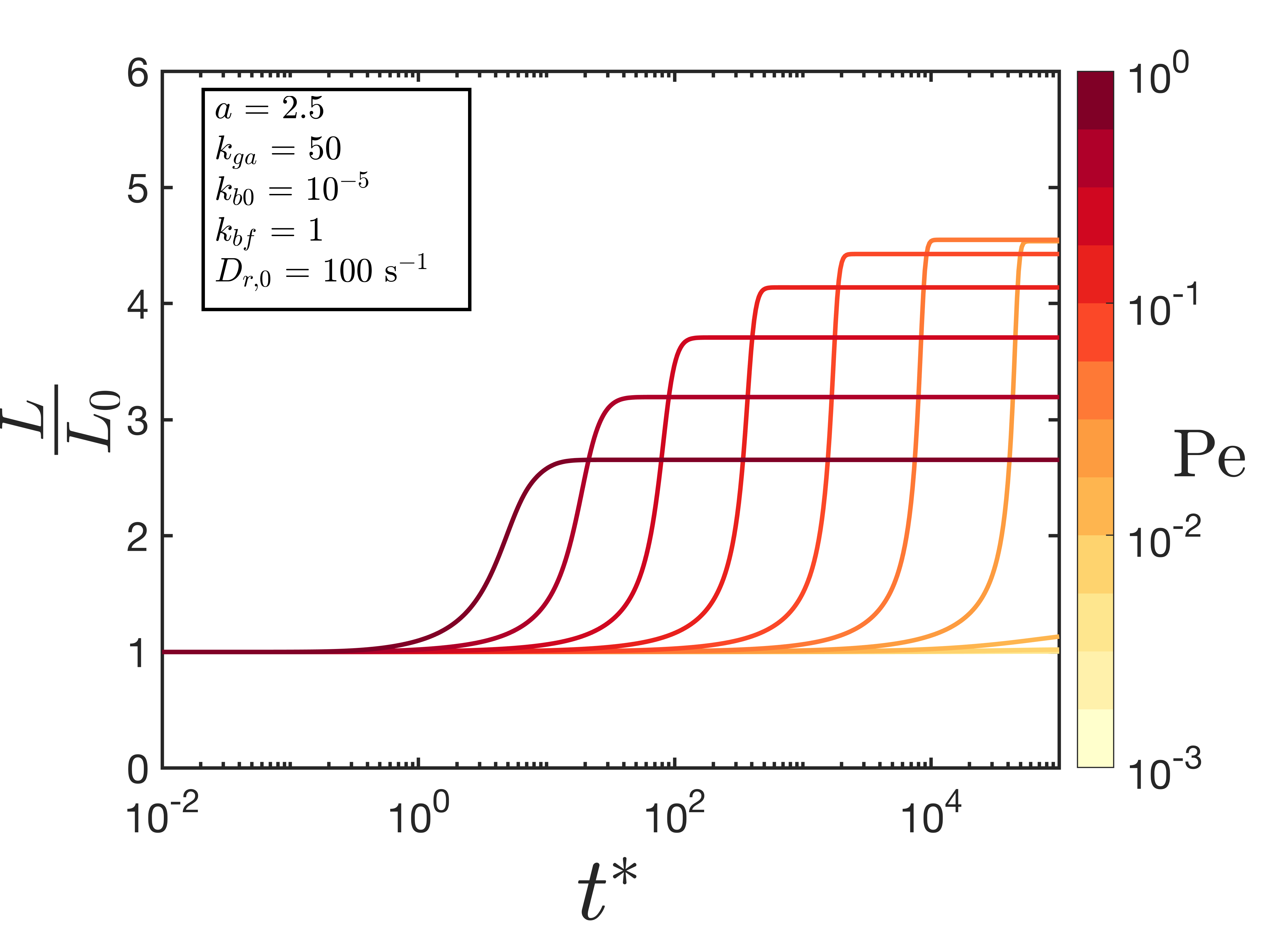}
        \label{fig:TransientShearL1}
    \end{subfigure}
    \begin{subfigure}{.5\textwidth}
    	\centering
        \subcaption{}
        \vspace{-3mm}
        \includegraphics[width=\linewidth]{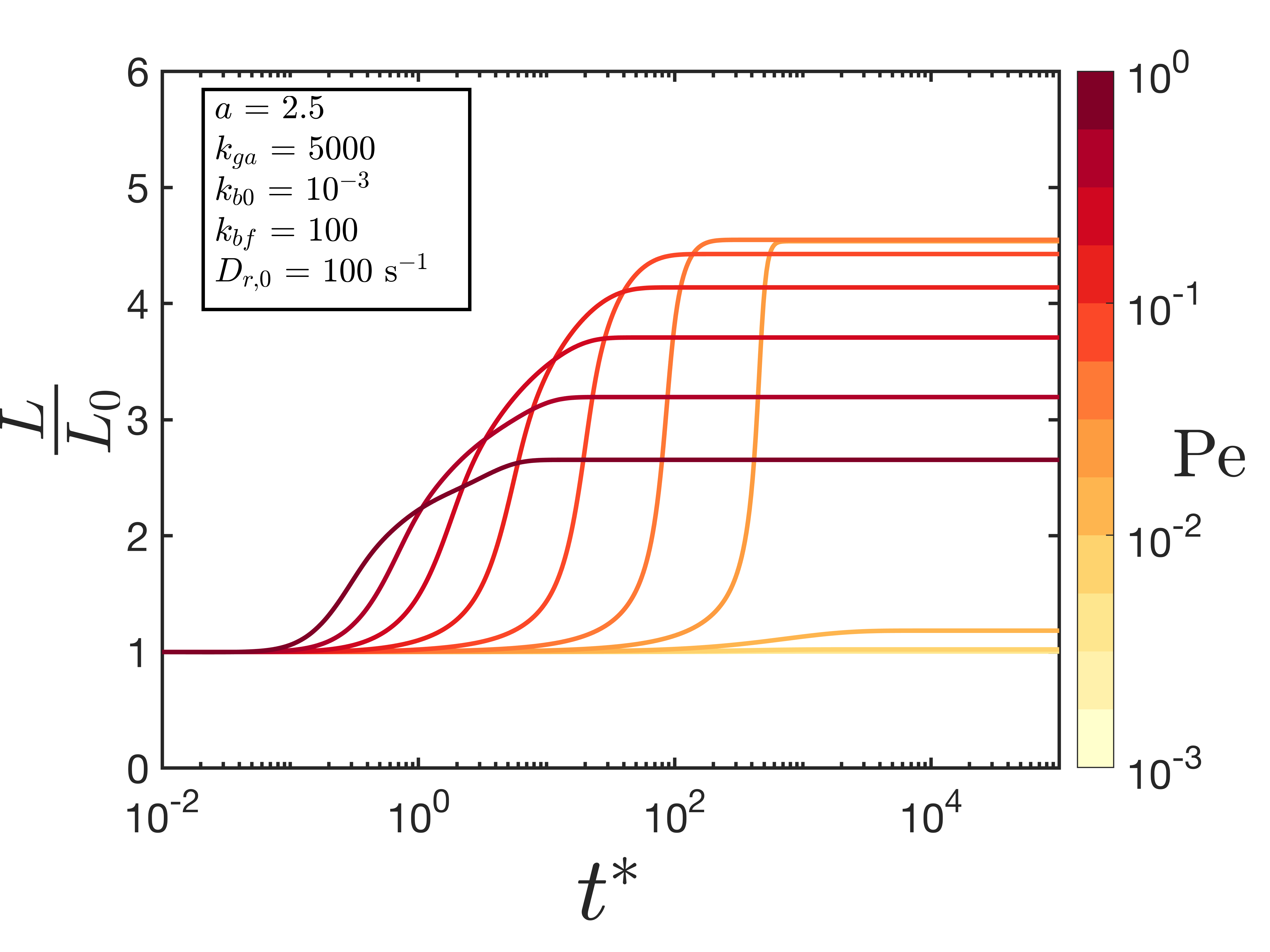}  
        \label{fig:TransientShearL2}
    \end{subfigure}
    \vspace{-6mm}
     \caption{Micellar contribution to the shear viscosity (normalized by zero-shear viscosity), ensemble average orientation components, and normalized length vs. dimensionless time in transient startup of steady shear flow for a range of applied \Peclet{} numbers. Both sets of parameters in \subref{fig:TransientShearVisc1}-\subref{fig:TransientShearL1} and \subref{fig:TransientShearVisc2}-\subref{fig:TransientShearL2} yield equivalent steady states but varying transient behavior.}
    \label{fig:TransientParameterSpace}
\end{figure*}

\Cref{fig:TransientShearVisc1,fig:TransientShearVisc2} show the time- and applied shear rate-dependent micellar contribution to viscosity as a function of dimensionless time for three decades of \Peclet{} numbers. Viscosity is normalized by the steady zero-shear viscosity, which importantly is not equal to the viscosity at $t=0$ (discussed in detail below). Darker lines corresponds to larger \Peclet{} numbers and the black dotted line corresponds to $\eta_m(t,\dot{\gamma})/\eta_{m,0} = 1$. We see that there are distinctly two thickening regimes present; the first, which is identical for all applied shear rates, occurs at very short times and is due to near-equilibrium alignment of rods. This slight thickening of the viscosity is caused by weak alignment of rods in the $S_{xy}$ direction, as seen in \cref{fig:TransientShearOrient1,fig:TransientShearOrient2}; this alignment is also seen in purely (i.e. non-reactive) Brownian rod suspensions and can be found analytically for startup of steady shear flow by solving for the near-equilibrium behavior of the rods. I.e.~to leading order at small $\Pe$, the transient viscosity is given by
\begin{equation}
    \eta_m(t^*) = \frac{n_0k_BT}{30D_{r,0}}\left(4-3\mathrm{e}^{-6t^*}\right).
\end{equation}
We can clearly see for  short times $\eta_m(t^*\ll 1) = \frac{n_0k_BT}{30D_{r,0}}$ which increases to $\eta_m(t^*\rightarrow\infty) = \frac{2n_0k_BT}{15D_{r,0}}$. We can also verify this by comparing \cref{fig:TransientShearOrient1} and \cref{fig:TransientShearL1}, in which growth in $S_{xy}$ can be seen occurring prior to elongation of micelles. 

The second thickening regime ($t^*\gtrsim 1$) can be attributed to elongation of the representative micellar length in tandem with rod alignment. We clearly see that the induction time $t_\text{ind}$, qualitatively defined here as the time for viscosity (and therefore stress) growth to occur, decreases with increasing shear rate, a trend that has been observed in experiments of WLM solutions \cite{Landazuri2016,Boltenhagen1997_2}.  In fact, as we will demonstrate later on, a distinctive power-law relationship between $t_\text{ind}$ and the applied shear rate $\dot{\gamma_0}$ can be observed, $t_\text{ind} \propto \dot{\gamma_0}^{-n}$, where $1 \lesssim n \lesssim 3$, which again agrees well with experimental observations in literature \cite{Landazuri2016}.

\begin{figure*}
	\centering
    \begin{subfigure}{.5\textwidth}
        \centering
        \subcaption{}
        \vspace{-3mm}
        \includegraphics[width=1\linewidth]{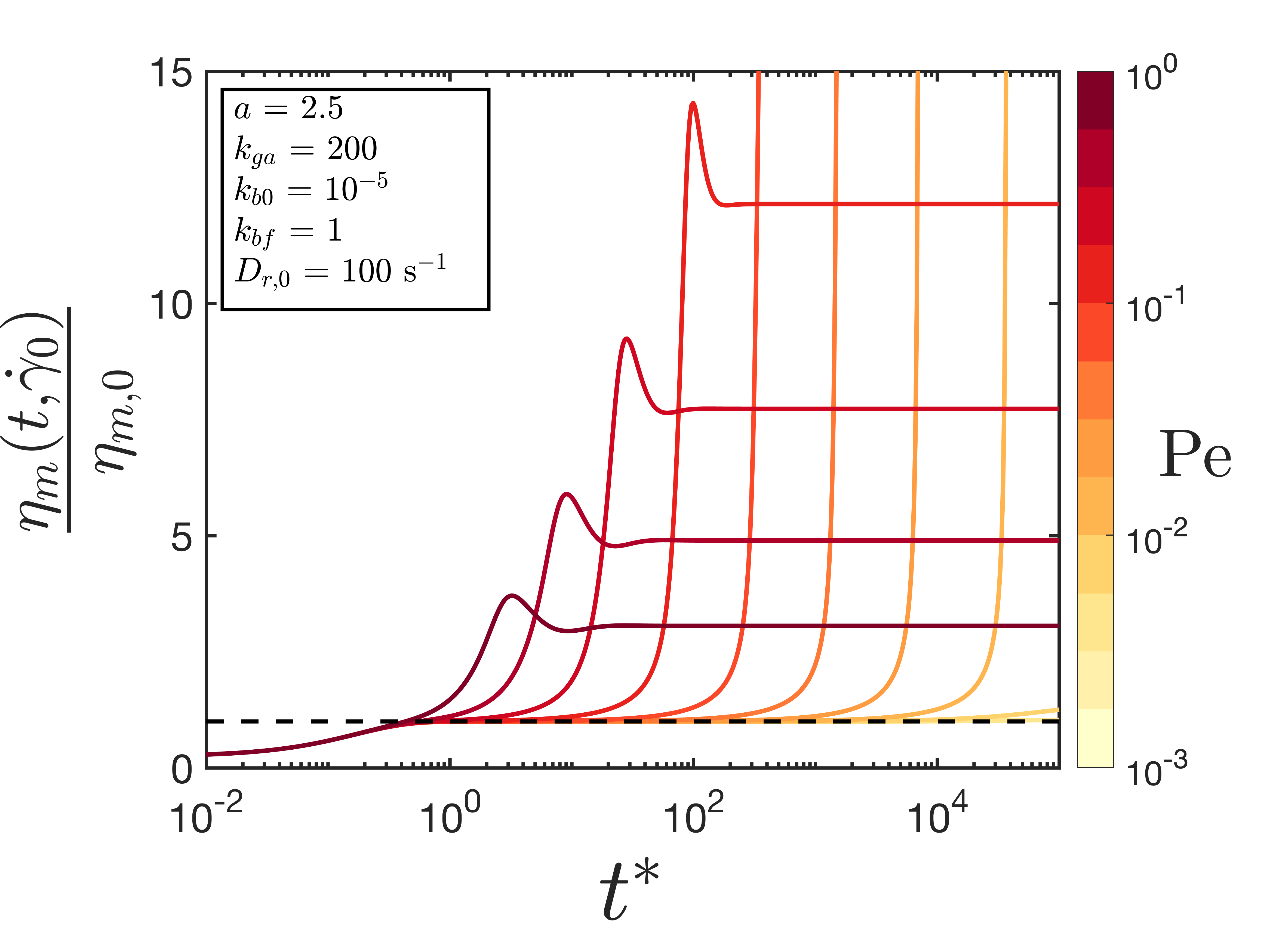}  
    	\label{fig:TransientShearVisc3}
    \end{subfigure}
    \vspace{-10mm}
    
    \begin{subfigure}{.5\textwidth}
        \centering
        \subcaption{}
        \vspace{-3mm}
        \includegraphics[width=1\linewidth]{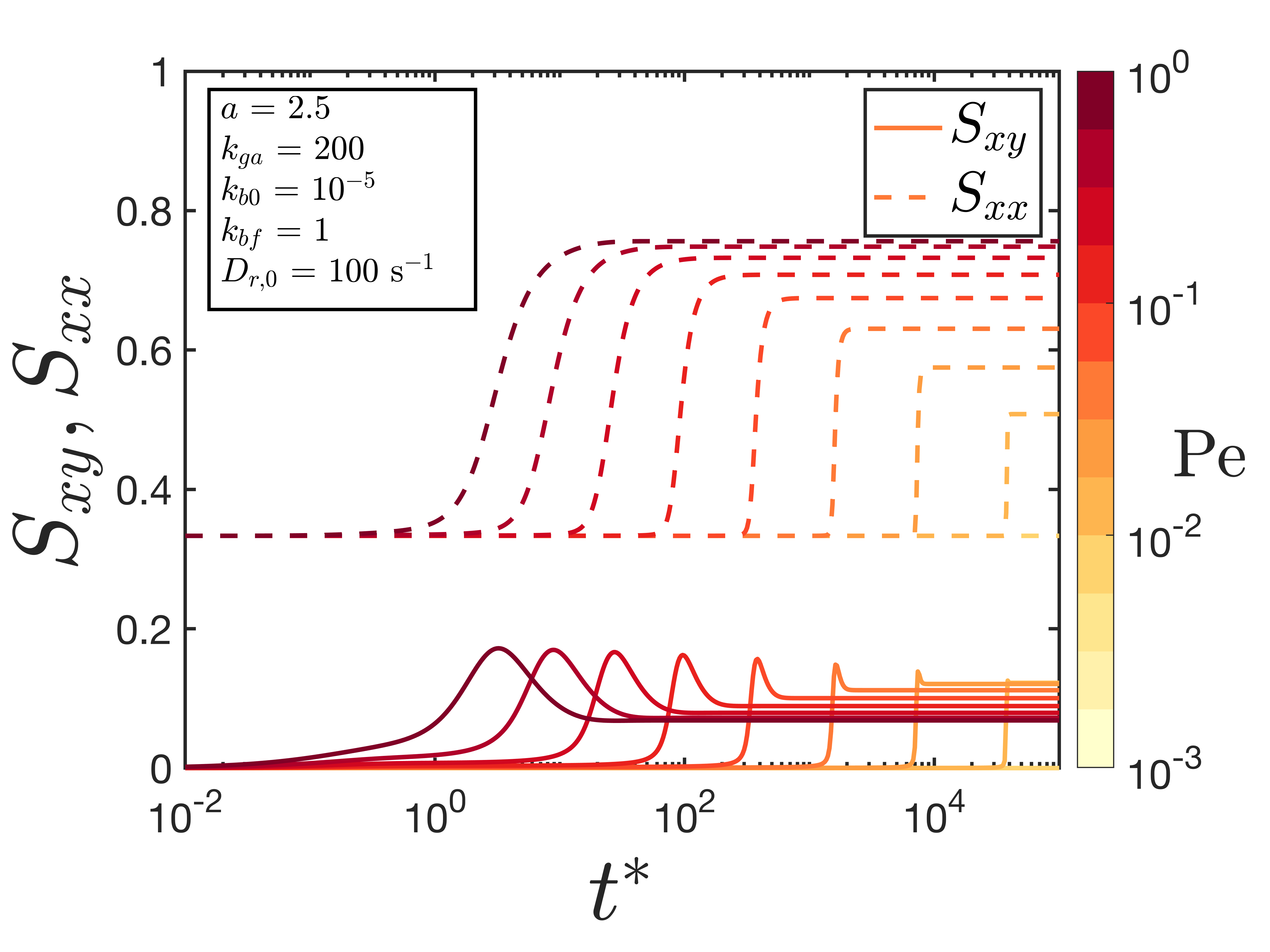} 
        \label{fig:TransientShearOrient3}
    \end{subfigure}
    \vspace{-10mm}
    
    \begin{subfigure}{.5\textwidth}
        \centering
        \subcaption{}
        \vspace{-3mm}
        \includegraphics[width=1\linewidth]{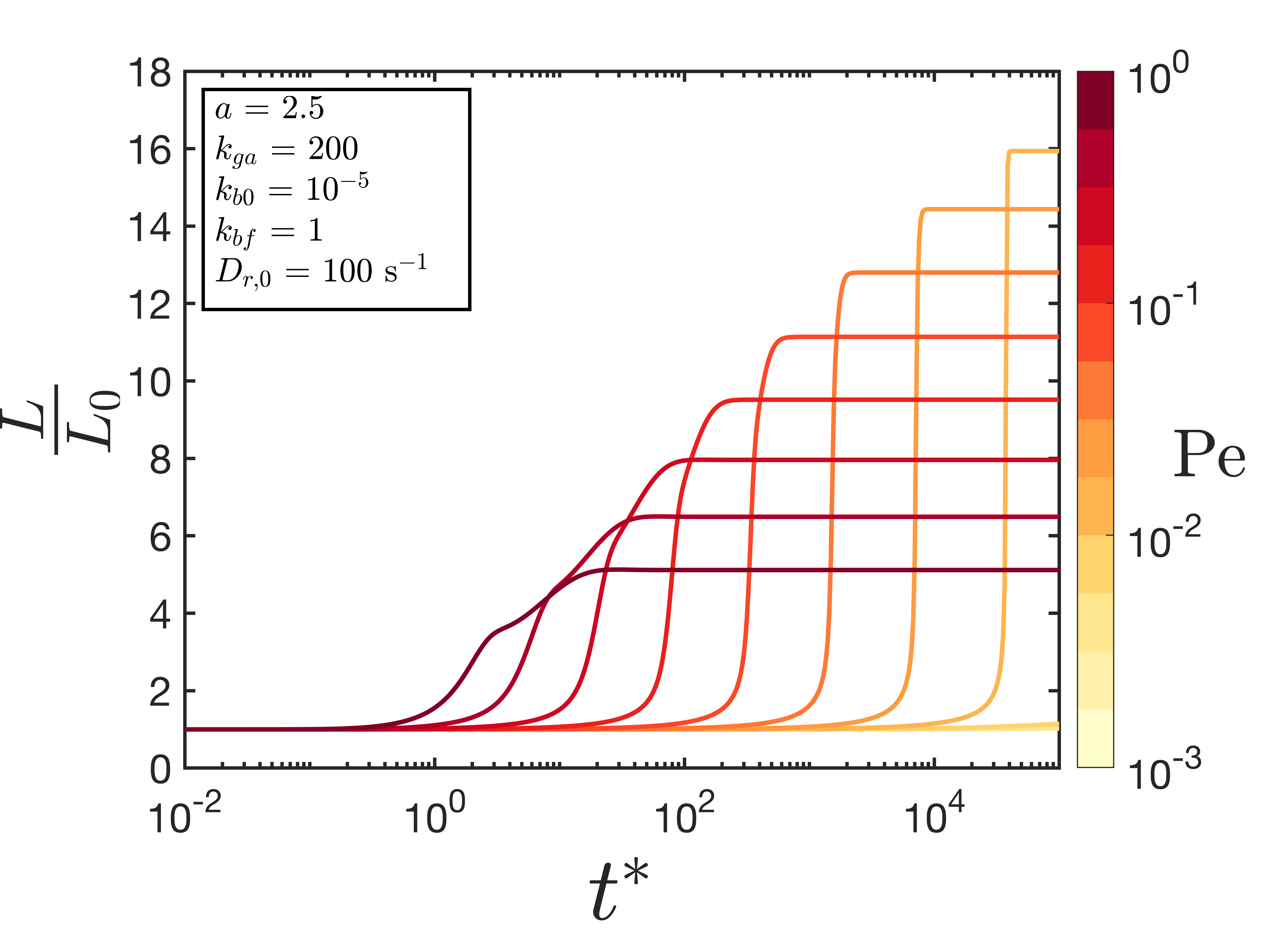}
        \label{fig:TransientShearL3}
    \end{subfigure}
    \vspace{-6mm}
     \caption{Micellar contribution to the shear viscosity (normalized by zero-shear viscosity), ensemble average orientation components, and normalized length vs. dimensionless time in transient startup of steady shear flow for a range of applied \Peclet{} numbers clearly demonstrating stress-overshoot phenomenon.}
    \label{fig:TransientOvershoot}
\end{figure*}

One feature of transient dynamics of dilute wormlike micelle solutions, particularly in startup of steady shear flow, is a viscosity (or stress) overshoot, in which the viscosity of the solution is observed to exceed its steady state viscosity before settling to the steady state value, oftentimes after a number of decaying oscillations \cite{Berret1997}. We can see slight instances of this overshoot in \cref{fig:TransientShearVisc2}, particularly for $\mathrm{Pe} > 10^{-1}$, but it is more clearly seen in \cref{fig:TransientShearVisc3} where we have increased the relative growth parameter $k_{ga}$ by a factor of four compared to \cref{fig:TransientParameterSpace}. By `relative' we mean in relation to $k_{b0}$ and $k_{bt}$. Comparing \cref{fig:TransientShearVisc3} with \cref{fig:TransientShearOrient3} and \cref{fig:TransientShearL3}, we see that this overshoot is solely a product of `over'-alignment in the $S_{xy}$ direction as there is no apparent overshoot in the micelle length, though notably in regions where $\eta_m$ and $S_{xy}$ are observed to overshoot there is a distinct change in the slope of $L^*$ vs. $t^*$. It is also interesting that the micelle length remains a monotonic function of time even following the viscosity overshoot, whereas $\eta_m$ and $S_{xy}$ do not. \RJHrevise{Stress overshoot also occurs in simple (i.e. constant length) Brownian rod suspensions and thus variations in micelle length are not necessary for an overshoot to occur. The monotonicity of the micelle length curve highlights the fact that micelle alignment and micelle elongation, although interrelated, are distinct phenomena.}

\begin{figure}
    \centering
    \begin{subfigure}{.5\textwidth}
        \centering
        \subcaption{}
        \includegraphics[width=.95\linewidth]{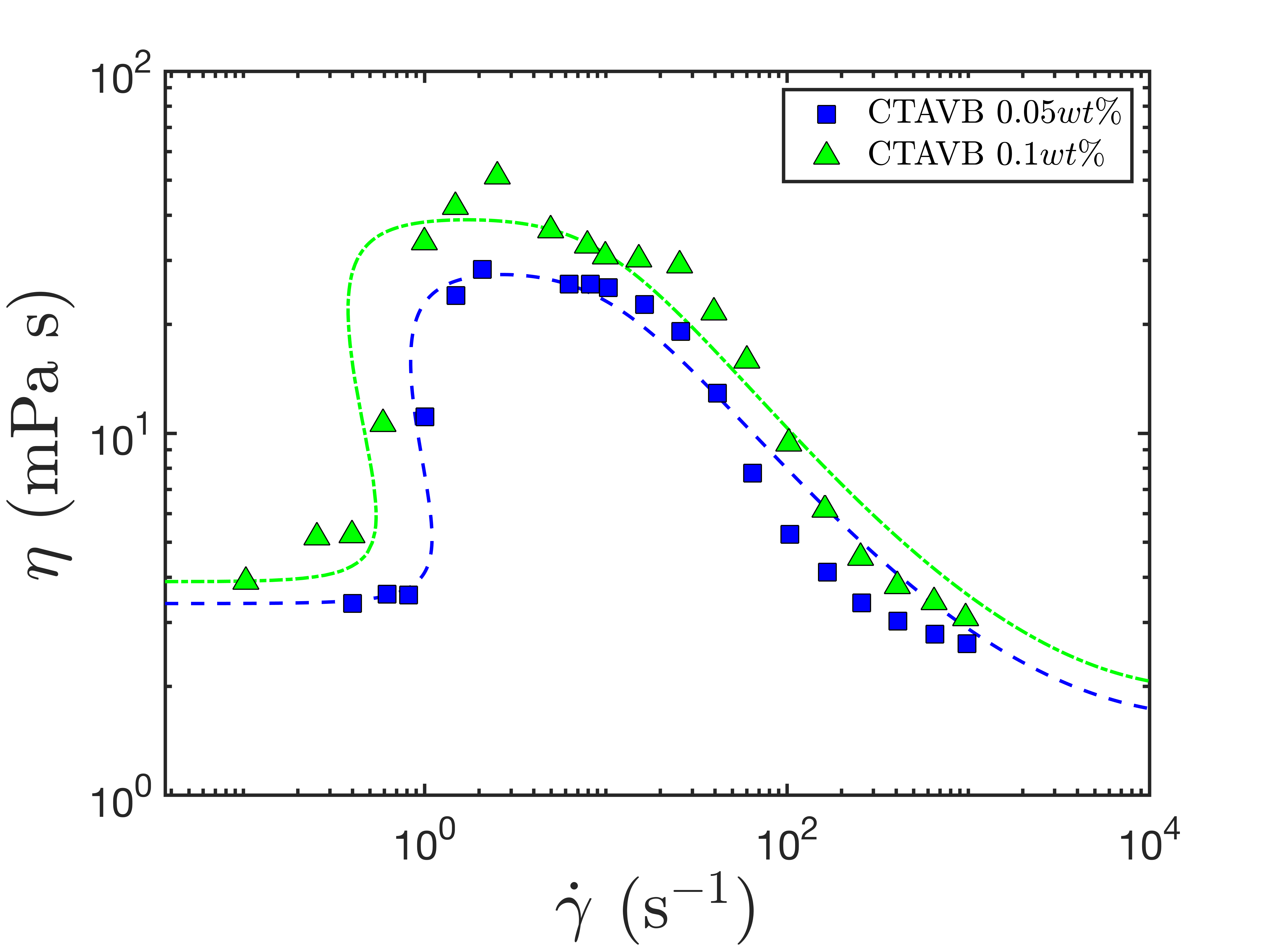}  
        \label{fig:BMP_visc}
        \subcaption{}
        \includegraphics[width=.95\linewidth]{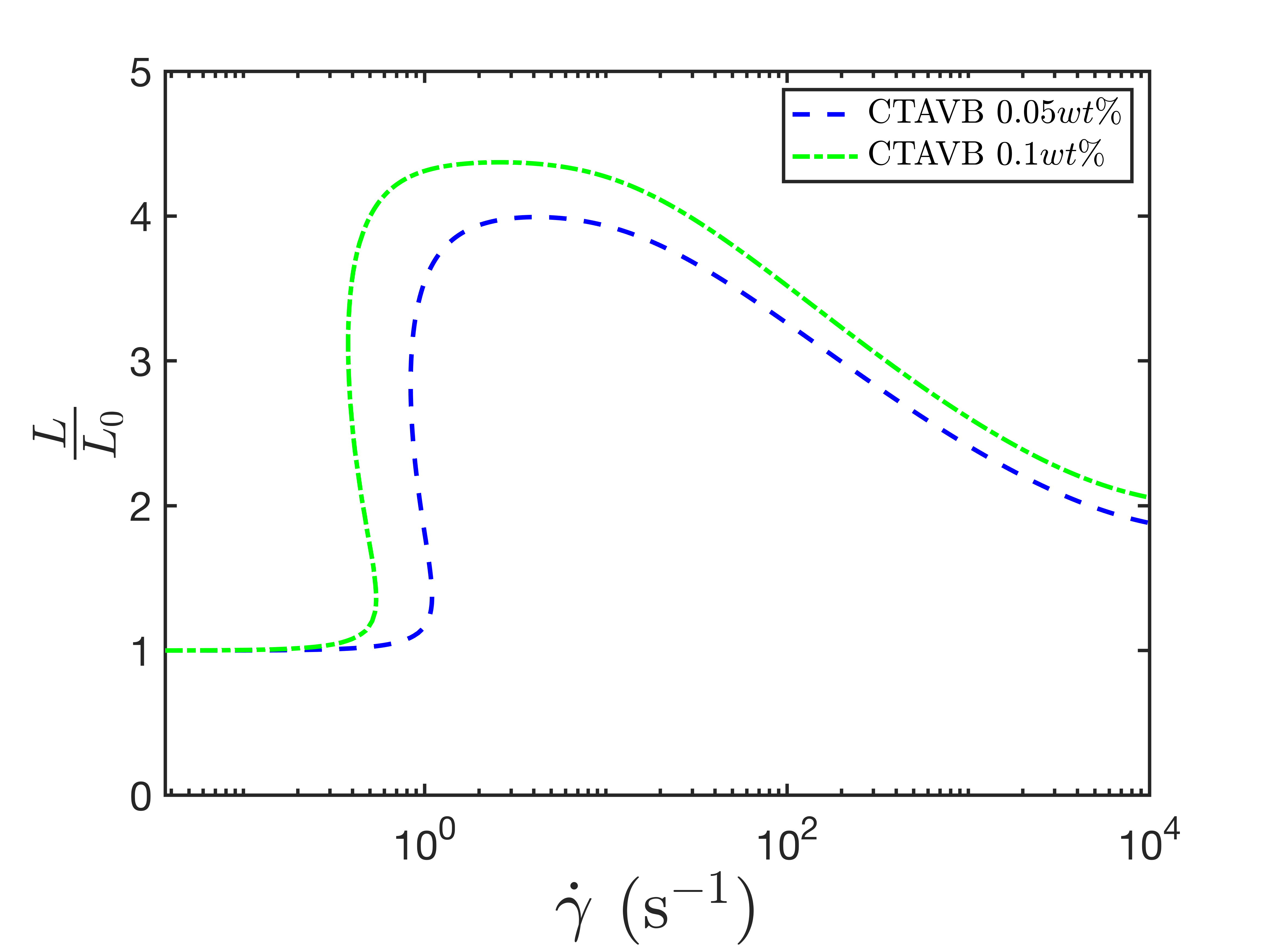}  
        \label{fig:BMP_L}
    \end{subfigure}
    \vspace{-3mm}
    \caption{Fits (lines) to experimental data (symbols) of \subref{fig:BMP_visc} shear viscosity vs. shear rate and \subref{fig:BMP_L} micelle length vs. shear rate. Experimental data corresponds to solutions of (green) $0.05 wt\%$ and $0.1 wt\%$ CTAVB in water obtained by Land\`azuri et al. \cite{Landazuri2016}. Corresponding parameter values are shown in \cref{table:BMP}.}
    \label{fig:BMP}
\end{figure}

\begin{figure}
    \centering
    \begin{subfigure}{.47\textwidth}
        \centering
        \subcaption{}
        \vspace{-3mm}
        \includegraphics[width=\linewidth]{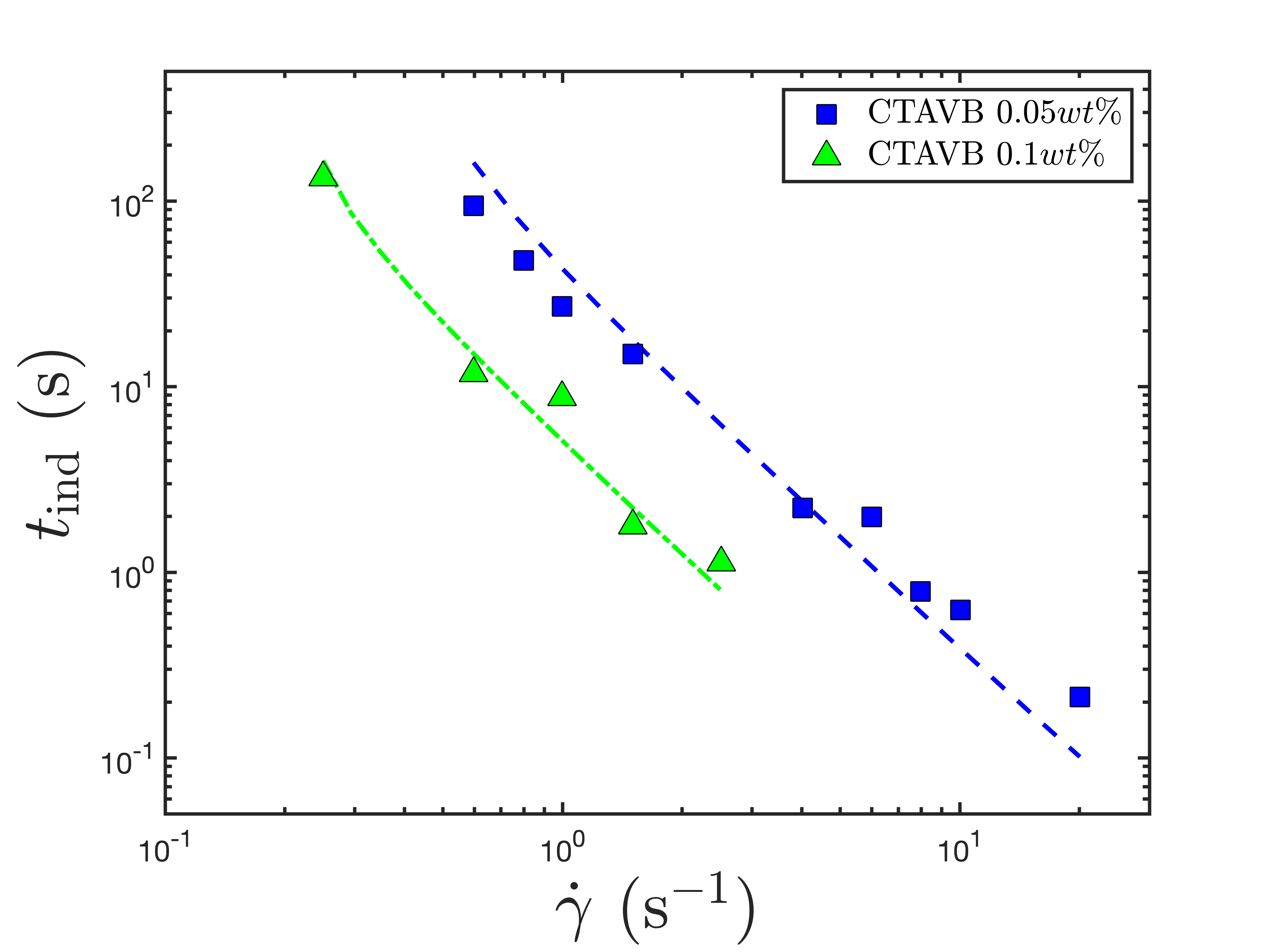}    
        \label{fig:BMP_tInd}
	\end{subfigure}
	\begin{subfigure}{.47\textwidth}
        \subcaption{}
        \vspace{-3mm}
        \includegraphics[width=\linewidth]{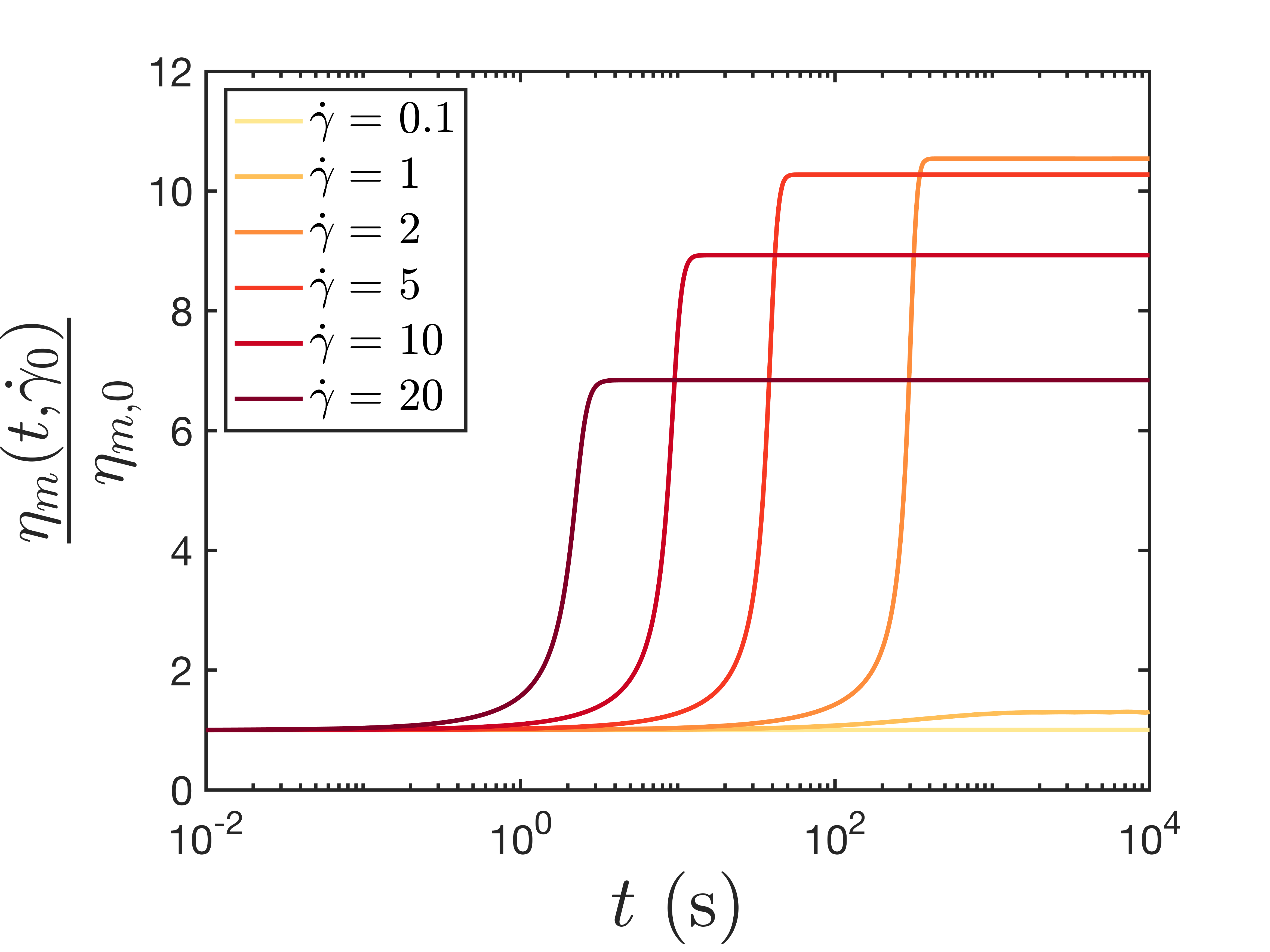}  
        \label{fig:BMP_trans}
    \end{subfigure}
    \vspace{-6mm}
    \caption{\RJHrevise{\subref{fig:BMP_tInd} Fits (lines) to experimental data of transient shear induction time vs. shear rate  for solutions of $0.05 wt\%$ and $0.1 wt\%$ CTAVB in water and \subref{fig:BMP_trans} corresponding transient viscosity growth vs. time response for the $0.05 wt\%$ solution (blue). Experimental data was obtained by Land\`azuri et al. \cite{Landazuri2016}. Corresponding parameter values are shown in \cref{table:BMP}.}}
    \label{fig:BMP_transient}
\end{figure}

We now turn from general features of the model to specific comparisons with experimental data that provide both steady and transient results. \Cref{fig:BMP} shows the \subref{fig:BMP_visc} steady state shear viscosity and \subref{fig:BMP_L} micelle length as a function of shear rate for $0.05 wt\%$ and $0.1 wt\%$ CTAVB in water solutions obtained by Land\`azuri et al. \cite{Landazuri2016}. Fits (lines) using the reformulated RRM are shown for steady shear viscosity. \Cref{table:BMP} shows the dimensional model parameters. We see that the model is well-suited for capturing steady shear viscosity and that predictions for length elongation are about $3.5$ times the equilibrium length. \Cref{fig:BMP_tInd} shows the corresponding predictions for the induction time ($t_\text{ind}$) as a function of applied shear rate for the same data sets \cite{Landazuri2016}. Fits are shown as lines. Note that we have taken the induction time to be defined as the time for stress growth to occur \textit{after} the viscosity increases to the steady state zero-shear viscosity, otherwise all curves would yield the same $t_\text{ind}$. As discussed previously, we can see a strong power-law relationship in \cref{fig:BMP_tInd} in which $t_\text{ind} \propto \dot{\gamma_0}^{-n}$. We find for the $0.05 wt\%$ CTAVB solution that $n = 2.05$ and for the $0.1 wt\%$ we find $n = 2.20$. The agreement between model predictions and experimental data for both steady and transient flows is quite good, and underscores the ability of our reformulated model to capture the dynamics of dilute wormlike micelle solutions. \RJHrevise{\Cref{fig:BMP_trans} shows the RRM-R predictions for transient micellar viscosity growth as a function of time which were used to generate the $t_\text{ind}$ vs. applied shear rate plot (\cref{fig:BMP_tInd}); explicit experimental data of viscosity (and stress) growth as a function of time were not available for this data set.}


\begin{table}
    \centering
    \caption{RRM-R parameters for experimental data (\cref{fig:BMP,fig:BMP_transient}) of CTAVB solutions \cite{Landazuri2016}.}
    \vspace{-3mm}
    \begin{tabularx}{0.6\textwidth} { 
         >{\raggedright\arraybackslash}X 
         >{\raggedright\arraybackslash}X 
         >{\raggedright\arraybackslash}X }
     \hline
     Composition & CTAVB & CTAVB \\
     $c$ [ppm] & 500 & 1000 \\ 
     \hline
     $k_{b0}$ [s$^{-1}$] & $1.9 \times 10^{-3}$ & $3.6 \times 10^{-3}$ \\ 
     $k_{ga}$ [$\mu\mathrm{m}^{7}$] & $4.3 \times 10^{-6}$ & $1.9 \times 10^{-5}$ \\ 
     $k_{bt}$ [m s$^{-1}$] & $2.2 \times 10^{-4}$ & $2.3 \times 10^{-3}$ \\ 
     $a$ [nm] & $2.0$ & $2.5$ \\ 
     $D_{r,0}$ [s$^{-1}$] & 120 & 140 \\ 
     \hline
    \end{tabularx}
    \label{table:BMP}
\end{table}

\subsection{Uniaxial extension: experimental comparison}
\label{sec:SteadyExtensional}
\begin{figure}
    \centering
		\vspace{-3mm}
        \includegraphics[width=.5\linewidth]{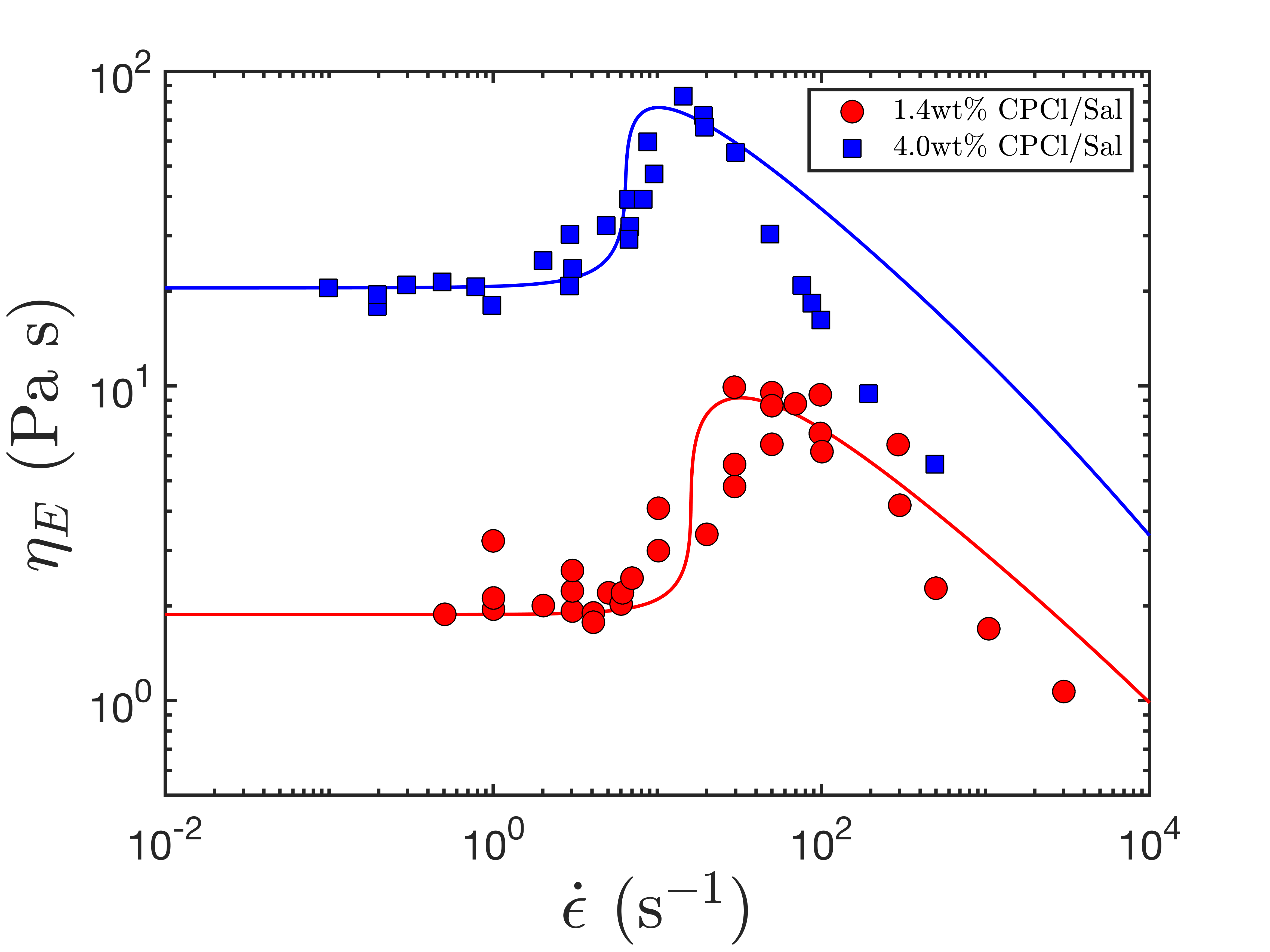}  
        
        \vspace{-3mm}
	    \caption{Fits (lines) to experimental data of extensional viscosity vs. strain rate for dilute (red) and semi-dilute (blue) CPyCl/NaSal solutions undergoing steady uniaxial extensional flow \cite{Walker1996}. Corresponding parameter values are shown in \cref{table:Walker}.}
	    \label{fig:Walker_visc}
\end{figure}

Having verified that our model is well-equipped to capture experimental results of dilute WLM solutions in both steady and transient shear flow, we turn our attention to extensional flows. Turbulent flows are dominated by extension and thus accurate prediction of extensional flow behavior is crucial if we aim to capture the mechanisms associated with surfactant-additive drag reduction in these flows \cite{Graham2014}. There are limited results of steady extensional flows of wormlike micelle solutions, primarily owing to the difficulty in performing these experiments. Walker and coworkers used an RFX opposed jet device to measure the extensional viscosity of semi-dilute concentrations of CPyCl/NaSal solutions in brine \cite{Walker1996}. These solutions are clearly not dilute, a fact that can easily be gleaned by noting that the zero-strain viscosities far exceed the viscosity of water. Fits to the experimental data are shown in \cref{fig:Walker_visc}, while model parameters are shown in \cref{table:Walker}. As we can from \cref{fig:Walker_visc}, the RRM-R is able to accurately predict the strain-hardening and -softening behavior observed in extensional flows of WLM solutions. Notably we have some difficulty in fitting the high-strain rate experimental data, which can likely be attributed to the non-dilute nature of the solutions; \RJHrevise{some fitting difficulty may also arise from the presence of pre-shear and/or micellar slippage effects that can often occur in these devices \cite{Rothstein2008}.} \RJHrevise{There has recently been significant interest in transient extensional experiments of dilute WLM solutions, notably capillary breakup extensional rheology (CaBER). We have performed preliminary work that suggests the RRM-R is well-suited for modeling CaBER and similar experiments, and we expect to communicate these findings in future work.}

\begin{table}
    \centering
    \caption{RRM-R parameters for experimental data (\cref{fig:Walker_visc}) of CPyCl/NaSal solutions \cite{Walker1996}.}
    \vspace{-3mm}
    \begin{tabularx}{0.53\textwidth} { 
         >{\raggedright\arraybackslash}X 
         >{\raggedright\arraybackslash}X 
         >{\raggedright\arraybackslash}X }
     \hline
     Composition & CPyCl & CPyCl \\
     $c$ [wt\%] & 1.4 & 4.0 \\ 
     \hline
     $k_{b0}$ [s$^{-1}$] & $0.50$ & $0.68$ \\ 
     $k_{ga}$ [$\mu\mathrm{m}^{7}$] & $7.0 \times 10^{-11}$ & $2.2 \times 10^{-11}$ \\ 
     $k_{bt}$ [m s$^{-1}$] & $2.0 \times 10^{-4}$ & $2.3 \times 10^{-5}$ \\ 
     $a$ [nm] & $1.5$ & $2.5$ \\ 
     $D_{r,0}$ [s$^{-1}$] & 100 & 15 \\ 
     \hline
    \end{tabularx}
    \label{table:Walker}
\end{table}


\section{Conclusions}

We have presented a reformulation (RRM-R) of the reactive rod constitutive model (RRM) that treats dilute surfactant solutions forming wormlike micelles as a suspension of rigid Brownian rods undergoing reversible scission and growth in flow. The model couples equations governing the ensemble average orientation, stress, and length of micelles to produce a dynamic set of equations allowing for the collective micelle length to elongate and breakdown. This framework produces steady shear viscosity vs. shear rate curves that exhibit drastic shear-thickening and shear-thinning regimes. Fits with the RRM-R to experimental data yields excellent agreement. The model depends on four dimensionless parameters describing: the spontaneous combination and breakdown of micelles ($k_{b0}^*$), growth due to alignment and collision of micelles ($k_{ga}^*$), and breakdown of micelles by tensile stresses ($k_{bt}^*$ and $a^*$). Certain combinations of parameters, particularly those where $k_{ga}^*$ is large and/or $k_{bt}^*$ and $a^*$ are small, produce reentrant (i.e. multivalued) steady state stress vs. shear rate curves, which is a necessary condition for a vorticity banding instability; this reentrant behavior indicates that the RRM-R is well-suited for studying this well-documented but poorly understood instability. Other parameter spaces, in which $k_{ga}^*$ is small and/or $k_{bt}^*$ and $a^*$ are large, do not undergo shear-thickening but rather show purely shear-thinning behavior. Although our model is intended to capture the behavior of dilute wormlike micelle solutions, we have shown that it is able to predict, at least partly, the behavior of semi-dilute solutions. 

The proposed model is also able to predict transient flow dynamics, in particular startup of steady shear flow, that well-aligns with observations seen in literature. The model predicts a power-law relationship between induction time and applied shear rate, which has been reported by numerous literature sources. The ability of the model to predict steady and transient flow behavior in tandem indicates that this constitutive formulation can be used in fluid-dynamics studies of complex flow behavior and instabilities. There is currently a limited understanding of dilute wormlike micelle solutions in turbulent flows and the manner in which they are able to achieve strong drag reduction; further, there has been limited research into the numerous instabilities, particularly the vorticity banding instability, present in dilute surfactant solutions. This model takes a large step towards uncovering these as-yet poorly understood phenomena.

\section{Acknowledgements}

This material is based on work supported by the National Science Foundation under grant number CBET-1803090.


\bibliography{mybibfile}

\end{document}